\pdfoutput=1

\documentclass[11pt]{article}

\usepackage{acl}
\usepackage{times}
\usepackage{latexsym}
\usepackage{amsmath,amssymb}
\usepackage{booktabs, colortbl, array}
\usepackage{comment}
\usepackage{listings}
\usepackage{varwidth}
\usepackage{bbm}
\usepackage{pifont}
\usepackage{mathbbol}
\usepackage[edges]{forest}
\usepackage{rotating}
\usepackage[utf8]{inputenc}
\usepackage{times}
\usepackage{pgfplots}
\usepackage{tikz}
\pgfplotsset{compat=newest}
\usepackage{latexsym}
\usepackage{marvosym}
\usepackage{arydshln}
\usepackage{rotating,tabularx,siunitx}
\definecolor{cYellow}{RGB}{255,255,3}
\definecolor{cBlue}{RGB}{69,123,157}
\definecolor{cRed}{RGB}{231,56,71}
\definecolor{dRed}{RGB}{255,20,20}
\definecolor{dOra}{RGB}{255,145,0}
\definecolor{dGreen}{RGB}{53, 94, 59}
\definecolor{cRed_1}{RGB}{191,30,46}
\definecolor{cGray}{RGB}{168,218,219}
\definecolor{dBlue}{RGB}{135, 206, 235}
\definecolor{cBlue_2}{RGB}{5,48,97}
\definecolor{cBlue_1}{RGB}{115,186,214}
\definecolor{cBlue_3}{RGB}{13,76,109}
\definecolor{cBlue_4}{RGB}{64,121,160}
\definecolor{cOrange}{RGB}{250,134,0}
\definecolor{cBlue_6}{RGB}{13,76,109}
\definecolor{cBlue_7}{RGB}{16,106,130}
\definecolor{cBlue_8}{RGB}{19,136,160}
\definecolor{cBlue_9}{RGB}{115,184,214}

\definecolor{above90}{RGB}{67,160,71} 
\definecolor{above80}{RGB}{13,76,109}
\definecolor{above70}{RGB}{215,175,24}
\definecolor{above60}{RGB}{244,81,30}
\definecolor{below60}{RGB}{90,90,90}

\definecolor{hidden-red}{RGB}{205, 44, 36}
\definecolor{hidden-blue}{RGB}{194,232,247}
\definecolor{hidden-orange}{RGB}{243,202,120}
\definecolor{hidden-green}{RGB}{34,139,34}
\definecolor{hidden-pink}{RGB}{245,235,237}
\definecolor{hidden-black}{RGB}{20,68,106}

\usepackage{float}
\usetikzlibrary{plotmarks}
\usepackage{graphicx}
\usepackage{subfigure}
\usepackage{caption}
\usepackage[ruled,vlined,linesnumbered]{algorithm2e}
\usepackage{mathtools}
\usepackage{booktabs}
\usepackage{multirow}
\usepackage{ragged2e}
\usepackage{hyperref}
\usepackage{multicol}
\usepackage{siunitx}
\usepackage{adjustbox}
\usepackage{pifont}
\usepackage{tabularx}
\usepackage{enumitem}
\usepackage{pythonhighlight}
\usepackage{array, makecell, multirow}
\newcolumntype{P}[1]{>{\centering\arraybackslash}p{#1}}
\usepackage{siunitx}
\usepackage{soul}
\usepackage[dvipsnames]{xcolor}
\usepackage{colortbl}
\sethlcolor{cYellow}
\usepackage{tcolorbox}
\usepackage{amsmath}
\usepackage{amssymb}
\usepackage{pifont}
\definecolor{coffee1}{RGB}{111,78,55}    
\definecolor{coffee2}{RGB}{212,163,115}  
\definecolor{coffee3}{RGB}{139,90,43}    
\definecolor{coffee4}{RGB}{222,184,135}  
\newcolumntype{R}[1]{>{\RaggedLeft\arraybackslash}p{#1}}
\newcolumntype{L}[1]{>{\RaggedRight\arraybackslash}p{#1}}
\tcbuselibrary{skins, breakable, theorems}

\usepackage[T1]{fontenc}

\usepackage[utf8]{inputenc}

\usetikzlibrary{shapes.geometric}

\newcommand{\change}[1]{\textcolor{black}{#1}}
\usepackage{microtype}
\usepackage{inconsolata}
\pgfplotsset{
/pgfplots/my legend/.style={
legend image code/.code={
    \node[star,star point ratio=2.25,rotate=36,minimum size=12pt,
          inner sep=0pt,draw=cBlue_4,solid,fill=cBlue_4,fill opacity=0.3] {};
   }
  }
}

\pgfplotsset{
/pgfplots/my medium/.style={
legend image code/.code={
    \node[fill opacity=0.7,text=cOrange] {\large $\odot$};
   }
  }
}
\pgfplotsset{
/pgfplots/my easy/.style={
legend image code/.code={
    \node[fill opacity=0.7,text=cBlue_3] {\large $\oplus$};
   }
  }
}

\pgfplotsset{
/pgfplots/my hard/.style={
legend image code/.code={
    \node[fill opacity=0.7,text=cRed_1] {\large $\otimes$};
   }
  }
}

\pgfdeclareplotmark{mystar}{
    \node[star,star point ratio=2.25,minimum size=12pt,rotate=36,
          inner sep=0pt,draw=cBlue_4,solid,fill=cBlue_4,fill opacity=0.3] {};
}
\pgfdeclareplotmark{my_easy}{
    \node[fill opacity=0.3,text=cBlue_3] {$\oplus$};
}
\pgfdeclareplotmark{my_medium}{
    \node[fill opacity=0.3,text=cOrange] {$\odot$};
}
\pgfdeclareplotmark{my_hard}{
    \node[fill opacity=0.3,text=cRed_1] {$\otimes$};
}
\usepackage{fdsymbol}

\title{AlphaQuanter: An End-to-End Tool-Augmented Agentic Reinforcement Learning Framework for Stock Trading}

\author{Zheye Deng$^{\clubsuit}$, Weixiang Yan$^{\vardiamondsuit}$, Changlong Yu$^{\clubsuit}$, Jiashu Wang$^{\clubsuit}$\\
  $^{\clubsuit}$Department of Computer Science and Engineering, HKUST, Hong Kong SAR, China\\$^{\vardiamondsuit}$University of California, Santa Barbara, CA, USA\\
  \texttt{zdengah@cse.ust.hk}\\
}

\begin{document}

\maketitle
\begin{abstract}

While Large Language Model (LLM) agents show promise in automated trading, they still face critical limitations. Prominent multi-agent frameworks often suffer from inefficiency, produce inconsistent signals, and lack the end-to-end optimization required to learn a coherent strategy from market feedback. To address this, we introduce \textbf{AlphaQuanter}, a single-agent framework that uses reinforcement learning (RL) to learn a dynamic policy over a transparent, tool-augmented decision workflow, which empowers a single agent to \textit{autonomously orchestrate tools} and \textit{proactively acquire information} on demand, establishing a transparent reasoning process. Extensive experiments demonstrate that AlphaQuanter achieves state-of-the-art performance on key financial metrics. \change{Besides, human evaluation shows the learned reasoning patterns reveal more faithful and coherent tool-usage behaviors, providing steps toward verifiable LLM-driven trading.} Our code and data can be found at \url{https://github.com/horizon-llm/AlphaQuanter}.

\end{abstract}

\section{Introduction}

The exploration of automated trading systems in modern financial markets is flourishing. Traditional machine learning methods (such as SVM, Random Forests, etc.)~\cite{rumelhart1986learning,cortes1995support, breiman2001random}  typically simplify the problem into discrete predictions of price direction at the next moment, making it difficult to effectively integrate multi-source heterogeneous trading signals. Although Deep Reinforcement Learning (DRL) can directly optimize decisions around long-term portfolio returns~\citep{NIPS1998_4e6cd952,wang2021deeptrader}, its black-box nature leads to trading decisions that \change{are difficult to trace back to concrete evidence.} Recently, Large Language Models (LLMs) have demonstrated tremendous potential in the field of financial trading~\citep{xiao2024tradingagents,DBLP:conf/kdd/ZhangZXSSQLZ0CZ24,wang2023alpha}. 

However, existing LLM-based attempts still face critical challenges. (1) First, \textit{the lack of tool orchestration and active information acquisition capabilities} makes it difficult for models to autonomously invoke and sequentially utilize external tools during the reasoning process, identify information gaps, and fill them on demand. (2) Second, \textit{\change{evidence-grounded decision traces are insufficient}}; current training paradigms are mostly black-box end-to-end optimization or offline answer fitting, lacking a \change{decision trajectory that can be consistently grounded in tool outputs and inspected step by step},
making it difficult to establish user trust and support regulatory audits. (3) Last but not least, \textit{prompt-based methods exhibit poor robustness}, are extremely sensitive to prompt engineering; meanwhile, multi-agent debate pipelines are often costly and redundant, and can further suffer from low decision efficiency, system fragility, and signal inconsistency. Overall, conducting reasoning under partially observable conditions, integrating heterogeneous signals, and executing actions with calibrated confidence remain core challenges that urgently need to be addressed.

To address these gaps, we propose \textbf{AlphaQuanter}, a single agent trading framework designed to enable \textit{informative}, \textit{\change{evidence-grounded}} and \textit{robust} trading decisions.
First, AlphaQuanter unifies the workflows into one ReAct-like agent~\cite{DBLP:conf/iclr/YaoZYDSN023} tailored for trading-oriented planning and reasoning.
We define several tools for various information sources and our framework starts from a guided plan followed by iterative tool use and information seeking as well as in-depth analysis.
Second, to further enhance decision-making capabilities and \change{promote faithful tool use with verifiable evidence},
we leverage reinforcement learning with verifiable rewards~\cite{guo2025deepseek,lambert2024tulu} to end-to-end optimize models that can selectively invoke useful tools and effectively gather supporting evidence.
We further curate high-quality outcome- and process-based reward signals to guide RL training across diverse actions.
This design eliminates the need for extensive prompt engineering across multiple agents, while \change{producing a more traceable decision trajectory and ensuring stable and efficient decision-making in practice.}
Finally, we evaluate our framework through backtesting protocols and \change{human evaluation.}
Our key contributions are summarized as follows:

\begin{figure*}[!t]
    \centering
    \includegraphics[width=\linewidth]{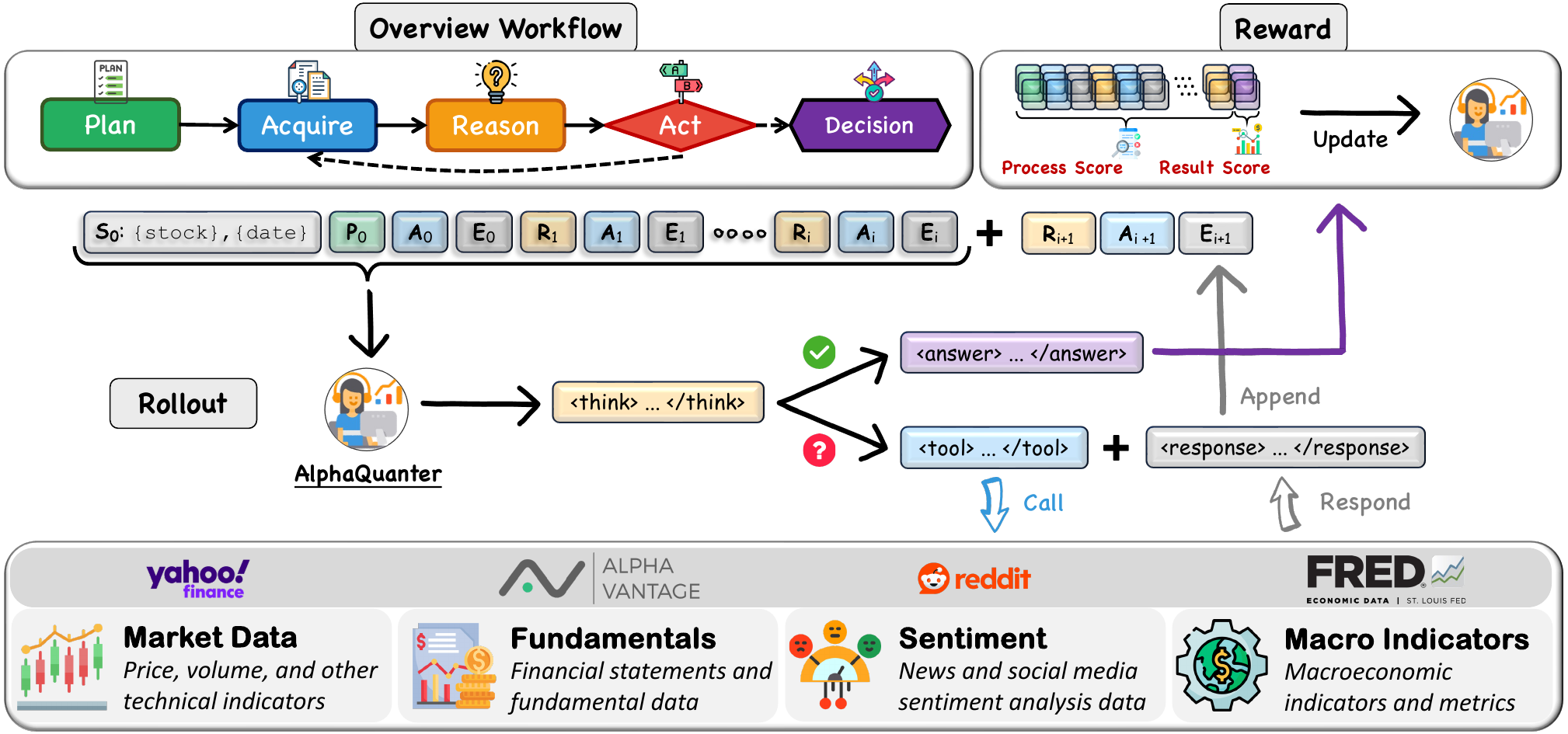}
    \caption{The overall architecture and workflow of AlphaQuanter. The central panel shows the agent's iterative rollout process. Starting from an initial state ($S_0$), the agent first forms an initial plan ($P_0$) before generating further reasoning traces ($R_{i+1}$) with \texttt{<think>} tag. In each step, it decides whether to continue acquiring information by executing a tool-based action ($A_{i+1}$) and receiving its environmental feedback ($E_{i+1}$), or to conclude by outputting a final decision with an \texttt{<answer>} tag. Throughout this process, the agent can query multi-dimensional financial data sources (\textit{bottom} panel, Section~\ref{sec:info_sources}). Once a decision is made, the entire trajectory will be evaluated to compute a reward (\textit{top-right} panel, Section~\ref{sec:score}), which updates the agent's policy. The overall workflow (\textit{top-left} panel, Section~\ref{sec:cognitive_workflow}) is designed to mimic a human trader's cognitive process of reasoning and acquiring data on demand.} 
    \label{fig:intro}
    \vspace{-10pt}
\end{figure*}

\begin{itemize}[leftmargin=10pt,itemsep=0pt]
    \item We propose a novel single-agent framework with effective reasoning chains that ensure both decision consistency and \change{traceable tool-use trajectories grounded in external evidence.}
    \item We design an end-to-end RL approach that trains the agent to actively acquire useful information and select supporting evidence for in-depth analysis. It directly optimizes the entire decision-making process for long-term profitability.
    \item Our empirical evaluations demonstrate that AlphaQuanter not only achieves state-of-the-art performance on key financial metrics \change{but also improves decision-trace faithfulness, exhibiting coherent, evidence-grounded} decision patterns that can be analyzed by human experts.
\end{itemize}

\section{Related Work}
Early approaches use traditional machine learning methods, such as SVM and random forest, to frame the task as a simple price direction classification~\cite{rumelhart1986learning,cortes1995support, breiman2001random}, which has been proven insufficient due to oversimplification and poor generalization in trading environments~\cite{zhong2025advancements}.

\paragraph{Deep Reinforcement Learning} \citet{NIPS1998_4e6cd952} pioneered the application of deep reinforcement learning to stock trading, directly optimizing trading performance end-to-end and outperforming supervised learning in long-horizon S\&P 500 backtests. 
iRDPG~\citep{liu2020adaptive} integrates imitation learning under a partially observable Markov decision process framework, using expert behavior to stabilize the training process and improve robustness, but overly relies on existing strategies. DeepTrader~\citep{wang2021deeptrader} introduces macro states and risk-sensitive rewards, achieving dynamic adjustment of long-short positions and risk control. MTS~\citep{DBLP:journals/corr/abs-2503-04143} improves returns across multiple datasets through time-aware encoding, parallel short selling, and CVaR-based risk management. However, these methods belong to end-to-end black-box optimization, lacking necessary interpretability, and cannot integrate external signals such as news and fundamentals on demand.

\paragraph{LLMs-Based Trading Agents} 
\change{FinMem~\cite{DBLP:journals/tbd/YuLCJLSZK25} exemplifies single-agent LLM trading systems that use layered memory retrieval and an explicit reflect step to construct prompts from multi-horizon financial events.} TradingAgents~\citep{xiao2024tradingagents} introduces a multi-agent framework where role-playing LLM agents debate to reach a trading decision, while FinAgent~\citep{DBLP:conf/kdd/ZhangZXSSQLZ0CZ24} further integrates multimodal information with tool-enhancement components, achieving competitive results across multiple evaluation metrics. However, these approaches lack explicit coordination and constraint mechanisms, making debate-style decision processes yield inconsistent or conflicting signals and high sensitivity to prompt design. Alpha-GPT~\citep{wang2023alpha} adopts a human-in-the-loop paradigm that enables factor mining through natural-language interaction, but it is difficult to autonomously scale and automate in high-frequency trading environments.

\paragraph{LLM-Based Reinforcement Learning Optimization} Motivated by the recent success of DeepSeek-R1~\citep{guo2025deepseek}, growing work explores RL approaches to optimize LLMs for quantitative trading. FLAG-Trader~\citep{DBLP:conf/acl/XiongDWCLYPLSLH25} employs partially fine-tuned LLMs as policy networks, optimizing trading rewards through policy gradient methods. Trading-R1~\citep{xiao2025trading} constructs large-scale financial corpora and implements a three-stage curriculum learning framework that combines SFT with RL. However, both types of methods generally lack end-to-end simulation of real trading processes and autonomous exploration capabilities and have not yet endowed models with spontaneous perception of information gaps or proactive orchestration of external tools.

\section{Problem Definition}
To navigate a partially observable market within a single trading day, we model the agent's task as a tool-augmented Markov Decision Process (MDP), defined by the tuple $\langle \mathcal{S}, \mathcal{A}, \mathcal{T}, \mathcal{R} \rangle$. The central challenge within this framework is 
to learn a strategy that sequences tool use and the final action to maximize return.

\paragraph{State Space $\mathcal{S}$}  
A state $s\in \mathcal{S}$ captures the agent's accumulated information, represented as the tuple
${\small s=(\texttt{initial\_context}, \texttt{query\_history}, \texttt{query\_result})}$, 
where \texttt{initial\_context} includes basic metadata (e.g., stock symbol, date), \texttt{query\_history} records the tools invoked so far, and \texttt{query\_result} stores their corresponding outputs.

\paragraph{Action Space $\mathcal{A}$} The action space $\mathcal{A}$ comprises two distinct types. First, the agent can execute a \textit{query action} from $\mathcal{A}_{q}=\{f_1,f_2,\cdots,f_{|\mathcal{A}_{q}|}\}$ to actively gather information from four source categories (market data, fundamental indicators, sentiment analysis, and macroeconomic metrics), thereby updating its state. Detailed descriptions of the data sources are provided in Section~\ref{sec:info_sources}. Finally, the agent can execute a \textit{decision action} from $\mathcal{A}_{d}=\{\texttt{BUY}, \texttt{SELL}, \texttt{HOLD}\}$, which terminates the decision-making process.

\paragraph{Transition Dynamics $\mathcal{T}$}
The state transitions are deterministic. 
When the agent selects a \textit{query action} $a_t \in \mathcal{A}_q$ at time step $t$, the current state $s_t$ transitions to $s_{t+1}$ by appending the query and its result to \texttt{query\_history} and \texttt{query\_results}, respectively. When the agent selects a \textit{decision action} $a_t \in \mathcal{A}_d$, the episode terminates immediately.

\paragraph{Reward Function $\mathcal{R}$}
An episode yields a trajectory $\tau=\left(s_0,a_0,s_1,a_1,\cdots,s_T,a_T\right)$, a sequence of states and actions beginning at the initial state $s_0$ and terminating with the first \textit{decision action} $a_T$, where  all intermediate actions $a_0,a_1,\cdots, a_{T-1}$ are \textit{query actions}. The agent's objective is to learn a policy $\pi$ that maximizes the cumulative trajectory reward $R(\tau)=\sum_{t=0}^T\mathcal{R}(s_t,a_t)$.
The step-wise reward function $\mathcal{R}: \mathcal{S}\times \mathcal{A} \rightarrow \mathbb{R}$ is designed to promote strategic and profitable decision-making:  rewarding \texttt{BUY} when the outlook is positive, \texttt{SELL} when negative, and \texttt{HOLD} when conditions are neutral or non-directional, while guiding tool use toward informative queries.

\section{AlphaQuanter} 
\label{sec:alpha_quanter}

\subsection{Cognitive Workflow}
\label{sec:cognitive_workflow}
Inspired by the ReAct paradigm~\cite{DBLP:conf/iclr/YaoZYDSN023}, \textbf{AlphaQuanter} interleaves reasoning traces with discrete \textit{actions}, as illustrated in Figure~\ref{fig:intro}. The workflow begins with an initial \textit{Plan} generation, followed by an iterative loop with three stages: (i) identify an information gap and \texttt{Acquire} new evidence via a tool call; (ii) \texttt{Reason} over the acquired evidence to update beliefs, and (iii) \texttt{Act} by either continuing the loop to gather more information or committing to a trading decision. This design enforces stepwise hypothesis testing while keeping evidence collection tightly coupled to reasoning. See Appendix~\ref{app:prompt_design} for the full prompt design.

\subsection{Reward Formulation}

\paragraph{Outcome Score}
\label{sec:score}
To train a robust, forward-looking agent under market noise, we encourage actions only on high-conviction signals, correctly classifying market states as strongly bullish, bearish, or neutral, while ignoring noise. We therefore smooth future returns by blending multiple horizons, akin to label smoothing~\cite{DBLP:conf/cvpr/SzegedyVISW16,DBLP:journals/corr/abs-2011-09607}. Specifically, we define the exponentially weighted forward return $r_t$ to filter short-lived fluctuations and emphasize the medium-term trajectory:
$r_t=\sum_{h=1}^{H}\omega_h\cdot \left({p_{t+h+1}}/{p_{t+1}}-1\right)$, where $p_t$ is the asset price on day $t$, $H$ is the maximum horizon, $\omega_h=\eta^h\big/\sum_{i=1}^H \eta^i$ is the normalized exponential weight, and $\eta \in (0,1)$ is the decay factor. Thresholding $r_t$ at $\theta$ yields the market regime, and we assign discrete rewards by action as specified in Table~\ref{tab:result_score}.

\begin{table}[!h]
    \centering
    \renewcommand{\arraystretch}{1.1}
    \scalebox{0.7}{
    \begin{tabular}{L{4cm}P{1.6cm}P{1.6cm}P{1.6cm}}
    \toprule
        \textbf{Future Market State} & $a_t=\texttt{BUY}$ & $a_t=\texttt{SELL}$ & $a_t=\texttt{HOLD}$  \\
    \midrule
    Highly Bullish ($r_t>\theta$) & \textbf{+1.0} & -1.0 & -0.75 \\
    Highly Bearish ($r_t<-\theta$) & -1.0 & \textbf{+1.0} & -0.75 \\
    Sideways ($|r_t|\leq \theta$) & -0.5 & -0.5 & \textbf{+1.0}\\
    \bottomrule
    \end{tabular}
    }
    \small
    \vspace{-8pt}
    \caption{Discrete reward structure for $\mathcal{R}_\text{result}$.}
    \vspace{-8pt}
    \label{tab:result_score}
\end{table}

\noindent We adopt an asymmetric penalty scheme to provide a more informative learning signal: taking the opposite side of a strong trend (reward $-1.0$) is penalized more than failing to act on an opportunity (reward $-0.75$), nudging the policy toward risk-aware behavior consistent with professional trading practice.

\paragraph{Process Score}
The process score comprises a format score $\mathcal{R}_{\text{format}}$ and a tool score $\mathcal{R}_{\text{tool}}$. The format score regulates the length of the reasoning trace: given a target token interval $[\min_\text{token}, \max_\text{token}]$, outputs outside this interval incur penalties, encouraging sufficiency without verbosity. The tool-use score governs acquisition efficiency by penalizing a total number of tool calls outside $[\min_{\text{tool}}, \max_{\text{tool}}]$. It further discourages the degenerate \emph{collect-then-conclude} pattern, acquiring all data in a single round and immediately producing a final answer, which can cause training to collapse. In addition, malformed tool calls that violate the function signature (e.g., missing or incorrect arguments) incur additional penalties. \change{See Appendix~\ref{app:process_score_computation} for the process-score pseudocode.}

\noindent In conclusion, we define the total reward $\mathcal{R}=\alpha\mathcal{R}_\text{result}+\mathcal{R}_\text{format}+\mathcal{R}_\text{tool}$, where the hyperparameter $\alpha$ places a greater emphasis on the outcome score, reflecting its primary importance.

\section{Evaluation}

\subsection{Backtesting Protocol}
While the policy is optimized for day-to-day decisions, its ultimate value is determined by the strategy's risk-adjusted performance over extended horizons~\cite{e5a1bb8f-41b7-35c6-95cd-8b366d3e99bc}. To evaluate this, we run backtests in which the daily-trained policy $\pi$ is applied sequentially across a historical period, generating a series of trades from which we measure the resulting portfolio performance.

\subsection{Portfolio State and Transition Dynamics}

\noindent We define the core variables of our portfolio in Table~\ref{tab:backtest_param}. 
The transition from the portfolio state $(h_t,c_t)$ to the new state $(h_{t+1},c_{t+1})$ is determined by the action $a_t$, as summarized in Table~\ref{tab:transitions}.

\begin{table}[h]
    \centering
    \renewcommand{\arraystretch}{1.1}
    \scalebox{0.68}{\begin{tabular}{P{1cm}m{9cm}}
    \toprule
        \textbf{Symbol} & \textbf{Description}  \\
    \midrule
        $p_i$ & Closing price of the asset on day $i$.  \\ 
        $h_i$ & Number of shares held at the end of day $i$. \\ 
        $c_i$ & Cash balance at the end of day $i$. \\ 
        $V_i$ & Total portfolio value at the end of day $i$. $\left(V_i=c_i+h_i\cdot p_i\right)$\\ 
        $\lambda$ & Transaction fee rate for BUY/SELL orders. \\ 
        $\kappa$ & Capital utilization ratio for BUY orders (slippage buffer). \\ 
    \bottomrule
    \end{tabular}
    }
\vspace{-8pt}
\small
    \caption{Backtesting Simulation Parameters.}
\vspace{-15pt}
    \label{tab:backtest_param}
\end{table}

\begin{table}[ht]
\centering
\renewcommand{\arraystretch}{1.1}

    \scalebox{0.7}{
\begin{tabular}{L{1cm}P{4cm}P{4cm}}
\toprule
\textbf{Action} & $\boldsymbol{h}_{\boldsymbol{t+1}}$ & $\boldsymbol{c}_{\boldsymbol{t+1}}$ \\
\midrule
\textsc{Buy}  &
$h_t + \left\lfloor \dfrac{\kappa\, c_t}{p_{t+1}} \right\rfloor$ &
$c_t - (1+\lambda)\, \left\lfloor \dfrac{\kappa\, c_t}{p_{t+1}} \right\rfloor p_{t+1}$ \\
\textsc{Sell} &
$0$ &
$c_t + (1-\lambda)\, h_t\, p_{t+1}$ \\
\textsc{Hold} &
$h_t$ &
$c_t$ \\
\bottomrule
\end{tabular}
}
\vspace{-8pt}
\caption{State transition rules for all actions.}
\vspace{-8pt}
\label{tab:transitions}
\end{table}

\subsection{Evaluation Metrics}
Following prior work~\cite{DBLP:conf/kdd/ZhangZXSSQLZ0CZ24,DBLP:conf/aaai/QinSZXW024,DBLP:conf/acl/XiongDWCLYPLSLH25}, we employ three widely used portfolio-level metrics as follows:

\noindent \textbf{Annualized Rate of Return (ARR)} measures profitability by annualizing total return:
\vspace{-5pt}
\begin{equation*}
\small
\text{ARR}=\left({V_T}/{V_0}\right)^{{252}/{T}}-1
\vspace{-5pt}
\end{equation*}
where $V_0$ and $V_T$ are initial and final portfolio values, $T$ is trading days, 252 is annual trading days.

\noindent \textbf{Sharpe Ratio (SR)} measures risk-adjusted performance: $\text{SR}=\bar{r}/{\sigma_r}$, where $r_t=(V_t-V_{t-1})/{V_{t-1}}$, $\bar r=\frac{1}{T}\sum_{t=1}^{T} r_t$, and $\sigma_r=\sqrt{\frac{1}{T-1}\sum_{t=1}^{T}\!\left(r_t-\bar r\right)^2}$. Higher SR indicates better risk-adjusted returns.

\noindent  \textbf{Maximum DrawDown (MDD)} measures the largest peak-to-trough decline:
\begin{equation*}
\small
    \text{MDD}=\max_{1\leq t\leq T}\left( \frac{\max_{1\leq s\leq t} V_s-V_t}{\max_{1\leq s \leq t}V_s}\right)
\end{equation*}
Lower MDD reflects better downside risk control.

\section{Experimental Setup} 
\label{sec:experiment}

\subsection{Dataset and Simulation Period}

\change{Following prior work~\cite{DBLP:conf/kdd/ZhangZXSSQLZ0CZ24, DBLP:journals/tbd/YuLCJLSZK25, DBLP:conf/acl/XiongDWCLYPLSLH25}, we select a widely used set of large, liquid stocks with frequent news activity, enabling fair comparison across methods.} We focus on five stocks: Alphabet Inc. (\textbf{GOOGL}); Microsoft Corporation (\textbf{MSFT}); Meta Platforms, Inc. (\textbf{META}); NVIDIA Corporation (\textbf{NVDA}); and Tesla, Inc. (\textbf{TSLA}). 
These firms are information-rich, providing rapidly evolving signals that stress iterative tool use and analysis.
We split data chronologically into non-overlapping train/valid/test sets, as shown in Table~\ref{tab:dataset_split}.
We also insert an approximately 30-trading-day gap between successive sets to prevent leakage and look-ahead bias.

\begin{table}[h]
    \centering
    \renewcommand{\arraystretch}{1.1}
    \scalebox{0.7}{
    \begin{tabular}{L{1cm}P{2cm}P{2cm}P{1.6cm}P{2.5cm}}
    \toprule
       \textbf{Set} & \textbf{Start Date} & \textbf{End Date} & \change{\textbf{Span}} & \textbf{\#Trading Days} \\
       \midrule
       Train & 2022-09-01 & 2024-03-30 & 19 months & 395 \\
       Valid & 2024-05-15 & 2024-11-14 & 6 months & 128 \\
       Test & 2025-01-01 & 2025-06-30 & \change{6 months} & 122 \\
    \bottomrule
    \end{tabular}
    }
    \caption{Dataset splits and trading-day counts.}
    \vspace{-7pt}
    \label{tab:dataset_split}
\end{table}

\begin{table*}[t]
    \centering
    \renewcommand{\arraystretch}{1.08}
    \scalebox{0.7}{
        \begin{tabular}{P{3.3cm}L{3.3cm}R{1.5cm}R{1.5cm}R{1.5cm}R{1.5cm}R{1.5cm}>{\columncolor{dBlue!20}}R{1.5cm}>{\columncolor{dBlue!20}}R{1.3cm}>{\columncolor{dBlue!20}}R{1.5cm}}
        \toprule
        \multirow{2}{*}{\textbf{Category}} & \multirow{2}{*}{\textbf{Model}}  & {\textbf{GOOGL}}  & {\textbf{META}} & \textbf{MSFT} & \textbf{NVDA} & \textbf{TSLA}  & \multicolumn{3}{>{\columncolor{dBlue!20}}c}{\textbf{Average}} \\
         \cmidrule(r){3-3} \cmidrule(r){4-4} \cmidrule(r){5-5} \cmidrule(r){6-6} \cmidrule{7-7} \cmidrule(r){8-10}
        & & \textit{ARR} ($\uparrow$) & \textit{ARR} ($\uparrow$)  & \textit{ARR} ($\uparrow$) & \textit{ARR} ($\uparrow$) & \textit{ARR} ($\uparrow$) & \textit{ARR} ($\uparrow$) & \textit{SR} ($\uparrow$) & \textit{MDD} ($\downarrow$)  \\
        \midrule
{\textbf{Market}} & B\&H & -14.49\% & 45.64\% & 36.80\% & 25.47\% & -28.91\% & 12.90\% & {{0.57}} & 31.13\% \\
\midrule
\multirow{2}{*}{{\textbf{Rule}}} & MACD & -3.17\% & 46.82\% & -9.58\% & -12.89\% & 22.77\% & 8.79\% & 0.44 & 21.24\% \\
& ZMR & -2.26\% & -0.98\% & 8.53\% & {{35.01\%}} & 16.74\% & 11.41\% & 0.46 & 20.86\% \\\midrule
\multirow{2}{*}{{\textbf{RL}}} & FinRL\textsubscript{A2C} & -21.22\% & 43.41\%  &  43.15\% & {{37.43\%}} & -35.10\% & 13.53\% & 0.51 & 33.07\% \\
& FinRL\textsubscript{PPO} & -19.77\% & 50.34\% & \textcolor{dOra}{\textbf{43.91\%}} & 18.56\% & -31.94\% & 12.22\% & 0.51 & 33.72\% \\
\midrule
\multirow{1}{*}{{\textbf{LM}}} & Chronos-2 & \textcolor{dRed}{\textbf{19.07\%}} & -12.61\% & 20.04\% & \textcolor{dOra}{\textbf{38.19\%}} & -17.66\% & 9.41\% & 0.34 & 24.34\% \\
\midrule
\multirow{2}{*}{{\textbf{LLM}}} & FinMem & -22.41\% & 46.25\% & 40.26\% & 26.71\% & -22.28\% & 13.71\% & 0.30 & 29.14\% \\
& TradingAgents &-14.95\% & 29.69\% & 38.62\% & -7.83\% & {{36.92\%}} & 16.49\% & 0.50 & 21.82\% \\
\midrule
\multirow{7}{*}{\makecell{\textbf{Multi-Agent}\\ \footnotesize({TradingAgents)}}} & Qwen2.5-3B & 1.73\% & 36.25\% & 40.89\% & -3.28\% & -76.98\% & -0.28\% & -0.13 & 20.95\% \\
& Qwen2.5-7B & \textcolor{dOra}{\textbf{9.33\%}} & 28.98\% & -4.50\% & -17.22\% & -9.11\% & 1.50\% & -0.08 & 6.43\% \\
& Qwen3-30B-A3B & -18.09\% & 1.36\% & 9.84\% & 10.22\% & -16.51\% & -2.64\% & 0.06 & 22.20\% \\
& DeepSeek-V3.1\textsubscript{\textit{685B}} & -12.43\% & -9.48\% & 14.13\% & -24.02\% & 0.00\% & -6.36\% & -0.26 & 12.49\% \\
& Kimi-K2\textsubscript{\textit{1T}} & -23.40\% & -9.52\% & 12.60\% & -8.33\% & 8.88\% & -3.95\% & -0.11 & 26.62\% \\
& GPT-4o-mini & -18.08\% & 0.73\% & 16.27\% & -5.38\% & 5.20\% & -0.25\% & -0.06 & 18.28\% \\
& GPT-4o & -14.95\% & 29.69\% & 38.62\% & -7.83\% & \textcolor{dOra}{\textbf{36.92\%}} & 16.49\% & 0.50 & 21.82\% \\
\midrule
\multirow{7}{*}{{\makecell{\textbf{Single-Agent}\\ \small({AlphaQuanter w/o RL)}}}}& Qwen2.5-3B & {{3.06\%}} & 23.08\% & 5.10\% & -7.43\% & -32.21\% & -1.68\% & 0.08 & 25.99\% \\
& Qwen2.5-7B & -22.42\% & 35.50\% & 17.55\% & 1.47\% & -9.63\% & 4.49\% & 0.16 & 28.96\% \\
& Qwen3-30B-A3B & -26.33\% & 32.86\% & 37.45\% & 29.61\% & -46.41\% & 5.44\% & 0.12 & 30.08\% \\
& DeepSeek-V3.1\textsubscript{\textit{685B}} & -25.15\% & 32.49\% & 25.45\% & 10.30\% & -1.21\% & 8.38\% & 0.24 & 30.70\% \\
& Kimi-K2\textsubscript{\textit{1T}} & -40.48\% & 25.83\% & -3.39\% & -3.27\% & 13.05\% & -1.65\% & 0.15 & 25.30\% \\
& GPT-4o-mini & -24.02\% & 44.42\% & {{43.42\%}} & 13.61\% & -43.71\% & 6.74\% & 0.25 & 26.78\% \\
& GPT-4o & -9.01\% & \textcolor{dRed}{\textbf{57.18\%}} & 19.39\% & 17.60\% & -38.04\% & 9.42\% & 0.25 & 28.27\% \\
\midrule
\multirow{2}{*}{{\makecell{\textbf{Single-Agent + RL}\\ \footnotesize({AlphaQuanter)}}}}&  AlphaQuanter-3B & -14.68\% & \textcolor{dOra}{\textbf{56.15\%}} & 9.82\% & 30.55\% & 33.33\% & \textcolor{dOra}{\textbf{23.03\%}} & 0.43 & 25.16\% \\
&  AlphaQuanter-7B & -2.52\% & 41.91\% & \textcolor{dRed}{\textbf{47.23\%}} & \textcolor{dRed}{\textbf{45.41\%}} & \textcolor{dRed}{\textbf{42.67\%}} & \textcolor{dRed}{\textbf{34.94\%}} & {{0.65}} & 24.93\% \\
\bottomrule
    \end{tabular}}
    \caption{Backtesting performance comparison of all methods over a \change{six-month} backtesting period. For the \textbf{\textit{ARR}} of each stock and the overall average, we mark the highest value in \textcolor{dRed}{\textbf{bold red}} and the second-highest in \textcolor{dOra}{\textbf{bold orange}}.} 
    \label{tab:main_result_122}
\end{table*}

\subsection{Information Sources} 
\label{sec:info_sources}

The tool integrates four primary data categories summarized below. See Appendix~\ref{appx:info_source} for details.

\paragraph{Market Data}
Daily price data (e.g., OHLCV) for each stock, along with a curated set of technical indicators grouped by function: trend (e.g., SMA, EMA), momentum (e.g., RSI, STOCH), volatility (e.g., BBANDS), and volume (e.g., OBV). These support technical analysis of market dynamics.

\paragraph{Fundamental Data}
Financial data from corporate filings, including income statements, balance sheets, cash flow statements, insider trading activity, dividend history, and earnings estimates. These support the assessment of company intrinsic value.

\paragraph{Sentiment Data}
Textual signals from financial news and social media platforms (e.g., Alpha Vantage, Reddit) to quantify market sentiment and investor psychology. These signals capture short-horizon sentiment and narrative shifts.

\paragraph{Macroeconomic Indicators}
Series capturing broad macroeconomic conditions and market-wide regimes, including inflation (e.g., CPI), interest rates (e.g., federal funds rate), and industry activity (e.g., commodity prices). These indicators provide the macroeconomic context for asset pricing.

\subsection{Implementation}

\paragraph{Baselines}
\change{We compare with seven categories of baselines: (1) \textbf{Market}: a passive \textit{buy and hold} strategy; (2) \textbf{Rule}: classic technical trading rules, including MACD and ZMR~\cite{DBLP:conf/kdd/ZhangZXSSQLZ0CZ24}; (3) \textbf{RL}: traditional deep RL baselines, including A2C and PPO, implemented in the FinRL framework~\cite{DBLP:journals/corr/abs-2011-09607} ; (4) \textbf{LM}: language modeling-based forecasting methods, represented by Chronos-2~\cite{DBLP:journals/corr/abs-2510-15821}; (5) \textbf{LLM}: LLM traders without RL optimization, including FinMem~\cite{DBLP:journals/tbd/YuLCJLSZK25} and TradingAgents~\cite{xiao2024tradingagents}; (6) \textbf{Multi-Agent}: the TradingAgents framework with different backbone LLMs: Qwen2.5-3B-Instruct, Qwen2.5-7B-Instruct~\cite{DBLP:journals/corr/abs-2412-15115}, Qwen3-30B-A3B-Instruct~\cite{DBLP:journals/corr/abs-2505-09388}, DeepSeek-V3.1~\cite{DBLP:journals/corr/abs-2412-19437}, Kimi-K2~\cite{DBLP:journals/corr/abs-2507-20534}, GPT-4o-mini, and GPT-4o~\cite{DBLP:journals/corr/abs-2410-21276}; (7) \textbf{Single-Agent}: an ablated AlphaQuanter that keeps the same tool access and prompting structure without RL training.}

\paragraph{Training Details}
We train the AlphaQuanter agents using Qwen2.5-3B-Instruct and Qwen2.5-7B-Instruct~\cite{DBLP:journals/corr/abs-2412-15115} as backbones, with the verl framework~\cite{DBLP:conf/eurosys/ShengZYWZZPL025}, optimizing the policy with the GRPO algorithm~\cite{shao2024deepseekmath}. At the inference time, we use deterministic decoding (temperature $=0$). For each configuration, we report the mean performance over three independent runs (distinct random seeds) to mitigate variance. All experiments are conducted on NVIDIA A100 GPUs (80GB). See Appendix~\ref{app:training_details} for detailed hyperparameters and training settings.

\section{Results and Analysis}

\subsection{Overall Performance Comparison}

To systematically evaluate tool-augmented and learning-based trading paradigms and demonstrate the effectiveness of our approach, we address three research questions. We report all metrics in Table~\ref{tab:main_result_122}.

\paragraph{RQ1: Single or multi-agent, which is better?}
We compare single-agent and multi-agent frameworks across multiple LLM backbones. The results show that, except for GPT-4o, the single-agent framework consistently outperforms the multi-agent framework on key metrics, particularly ARR. This supports our hypothesis that, for smaller-scale models, multi-agent \textit{debate} can inject noise or amplify hallucinations rather than yield complementary insights, ultimately degrading performance. These results provide clear justification for adopting a single-agent architecture in our approach.

\paragraph{RQ2: Is prompt-based reasoning sufficient for trading decisions?} 
We compare the strongest prompt-based baselines against the simple \textit{buy-and-hold} strategy. On average, all backbones except GPT-4o fail to beat the market method. 
We attribute the underperformance to difficulty learning actionable decision boundaries. Although the models can infer bullish or bearish sentiment, the prompt-based baseline does not reliably calibrate the decision threshold at which a signal should trigger \texttt{BUY} rather than \texttt{HOLD}. This exposes a fundamental limitation of current small-scale LLMs for trading and indicates that prompt-only reasoning is insufficient; agents should be explicitly trained to map high-dimensional market states to optimal trading actions.

\paragraph{RQ3: How effective is the AlphaQuanter?} 
We compare the fully trained \textbf{AlphaQuanter} against all baselines. Both the 3B and 7B variants significantly outperform the strongest baseline, with absolute ARR gains of 6.54\% and 18.45\%, respectively. \change{Notably, AlphaQuanter also surpasses deep RL agents that operate on numerical features and LLM baselines that directly ingest all available financial signals, highlighting the benefit of training a tool-using policy rather than a feature-only controller.}
Moreover, the 7B model is notably consistent, outperforming all baselines on three of the five stocks, showing that end-to-end RL training enables small LLMs to learn robust trading policies, including proactive tool use and information seeking, that even surpass powerful zero-shot setting such as GPT-4o. In short, the evidence indicates that specialized training paradigms may be more critical than model scale for achieving state-of-the-art performance in automated trading systems.

\subsection{Training Dynamics and Validation Performance}
\label{sec:tdvp}

\begin{figure}[t]
    \centering
    \small
    \includegraphics[width=\linewidth]{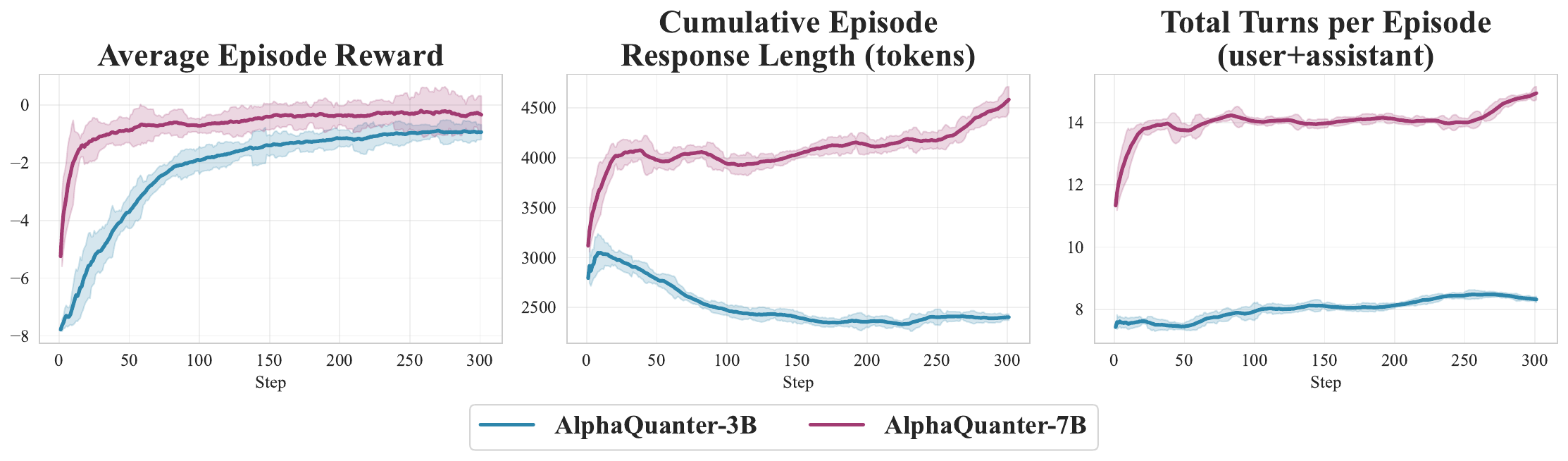}
    \caption{\change{Comparison of training dynamics for the AlphaQuanter-3B and -7B models.}}
    \label{fig:learning_dynamics}
\end{figure}

\begin{figure}[t]
    \centering
    \small
    \includegraphics[width=\linewidth]{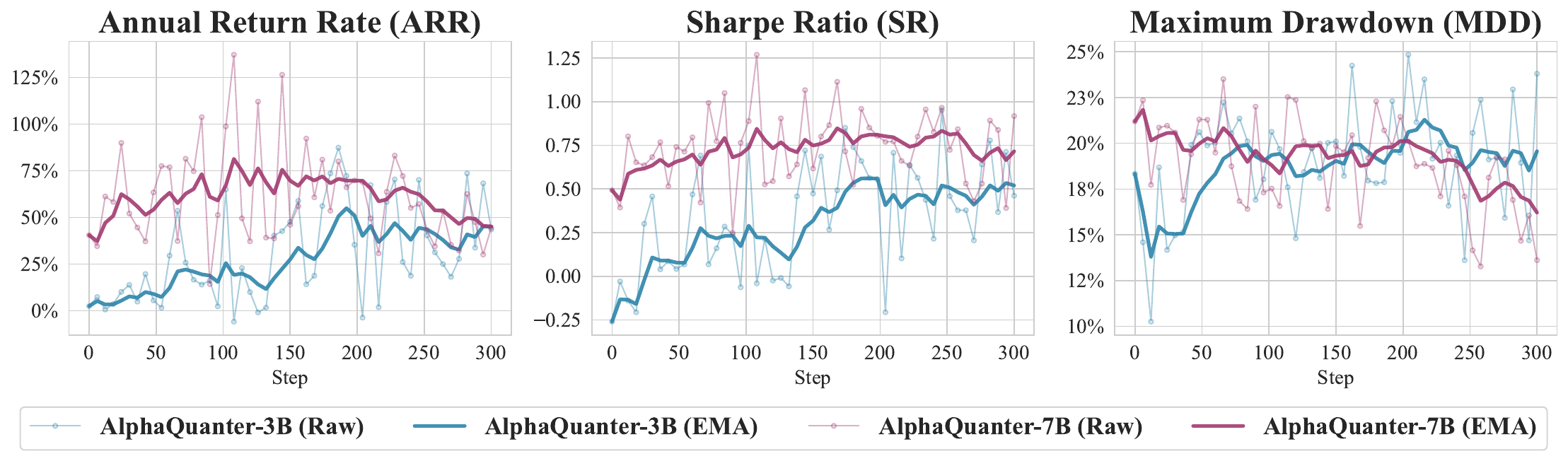}
    \vspace{-10pt}
    \caption{Comparison of key backtesting metrics for the AlphaQuanter-3B and -7B models on the validation set.}
    \label{fig:valid_perform}
    \vspace{-10pt}
\end{figure}

To capture the training process comprehensively, we track (1) \textit{training dynamics}~(Figure~\ref{fig:learning_dynamics}), reflecting reward signals and behavioral patterns during learning; and (2) \textit{validation performance}~(Figure~\ref{fig:valid_perform}), evaluating trading performance on unseen data. 
This joint analysis reveals how the agents' evolving behaviors translate into practical outcomes.

\paragraph{Training Dynamics}
The upward reward trajectory demonstrates effective learning from market interactions.
\change{Interpreting the behavioral metrics in Figure~\ref{fig:learning_dynamics}, response length is cumulative tokens per episode and turns count total user+assistant messages; these metrics reveal markedly different learning dynamics between the two models.} Both exhibit an initial volatile exploration phase, but their subsequent paths differ substantially. The 3B model transitions into a simplistic exploitation phase characterized by fewer tool calls and decreasing response length, suggesting premature convergence to a less robust policy. In contrast, the 7B model achieves stable exploitation around step 200 and subsequently enters a policy refinement phase, evidenced by increased response length and dialogue turns. This pattern indicates that the larger model explores more sophisticated reasoning chains and information-seeking strategies to extract marginal performance gains.

\paragraph{Validation Performance}
The validation metrics confirm the successful generalization of the learned policies to unseen data. For both 3B and 7B, ARR and SR exhibit clear upward trends that closely mirror the training reward curves. Notably, the 7B model shows a downward trend in MDD, indicating it has learned not only to maximize returns but also to effectively manage downside risk. Conversely, the 3B model's MDD oscillates with an upward bias, revealing its failure to internalize risk-aware trading behavior despite improving returns.

\paragraph{\change{Efficiency}}

\change{Beyond performance, we assess inference cost by measuring the average number of LLM-generated tokens per trading decision, aggregated over all trading days across the five datasets. Table~\ref{tab:efficiency} compares the multi-agent baseline (\textbf{MA}), a prompt-only single-agent (\textbf{SA}), and AlphaQuanter (\textbf{SA+RL}). The prompt-only single-agent is the most token-efficient, but its ARR is not the highest. AlphaQuanter achieves the best ARR while maintaining low token usage, corresponding to only \textbf{0.15$\times$} the cost of the multi-agent baseline.}
\begin{table}[t]
    \centering
    \scalebox{0.65}{
\begin{tabular}{L{3cm}P{1.5cm}P{2cm}P{1.6cm}P{1.6cm}}
\toprule
\textbf{Model} & \textbf{Setting} & \textbf{Avg. Tokens} & \textbf{Cost} & \textbf{ARR} \\
\midrule
TradingAgents-7B & MA  & 27.2K & 1.00$\times$ & \ \ 1.50\% \\
SingleAgent-7B & SA & \ \ 3.1K & \textbf{0.11$\times$} & \ \ 4.49\% \\
AlphaQuanter-7B  & SA+RL & \ \ 4.1K  & {0.15$\times$} &\textbf{34.94\%} \\
\bottomrule
\end{tabular}
}
\caption{\change{Token efficiency comparison for multi-agent (\textbf{MA}), prompt-only single-agent (\textbf{SA}), and the \textbf{RL}-trained single-agent AlphaQuanter.}}
\label{tab:efficiency}
\end{table}

\subsection{\change{Faithfulness Analysis}} \label{sec:faith_analysis}

\change{To quantify faithfulness of tool-augmented decision traces, we use three human-rated metrics: (1) \textit{\textbf{Alignment}} (whether stated information needs match executed tool calls), (2) \textit{Evidence \textbf{Grounding}} (whether key decision claims are supported by tool outputs), and (3) \textit{\textbf{Conciseness}} (whether the trace avoids redundant tool calls). We randomly sample 50 inputs and collect outputs from three methods (\textbf{MA}/\textbf{SA}/\textbf{SA+RL}). Three Ph.D. raters independently score each output on a $0-2$ scale for all metrics, and we assess inter-rater reliability using Krippendorff's $\alpha$~\cite{Hayes01042007} for ordinal ratings ($\alpha=0.78$). We report the averaged results in Table~\ref{tab:faithfulness}.
Overall, \textbf{SA+RL} (AlphaQuanter) achieves the best performance across all three metrics. In contrast, \textbf{MA} scores lowest on \textit{alignment} and \textit{conciseness} due to conflicting agent signals and redundant debate. Detailed analysis and scoring rubrics are provided in Appendix~\ref{app:detailed_faith_analysis}.}
\begin{table}[!t]
    \centering
    \scalebox{0.7}{
\begin{tabular}{L{1.4cm}P{2cm}P{2cm}P{2cm}P{1.4cm}}
\toprule
\textbf{Setting} & \textbf{Alignment} & \textbf{Grounding} & \textbf{Conciseness} & \textbf{Overall} \\
\midrule
MA & 1.20 & \underline{1.24} & 1.03 & 1.157 \\
SA & \underline{1.38} & 1.18 & \underline{1.32} & \underline{1.293} \\
SA+RL  & \textbf{1.70} & \textbf{1.28} & \textbf{1.68} & \textbf{1.557} \\
\bottomrule
\end{tabular}
}
\caption{Faithfulness analysis for TradingAgents (\textbf{MA}), prompt-only SingleAgent (\textbf{SA}), and the RL-trained single-agent AlphaQuanter (\textbf{SA+RL}).}
\label{tab:faithfulness}
\end{table}

\subsection{Tool Usage Patterns}

\begin{figure}[!t]
    \centering
    \includegraphics[width=\linewidth]{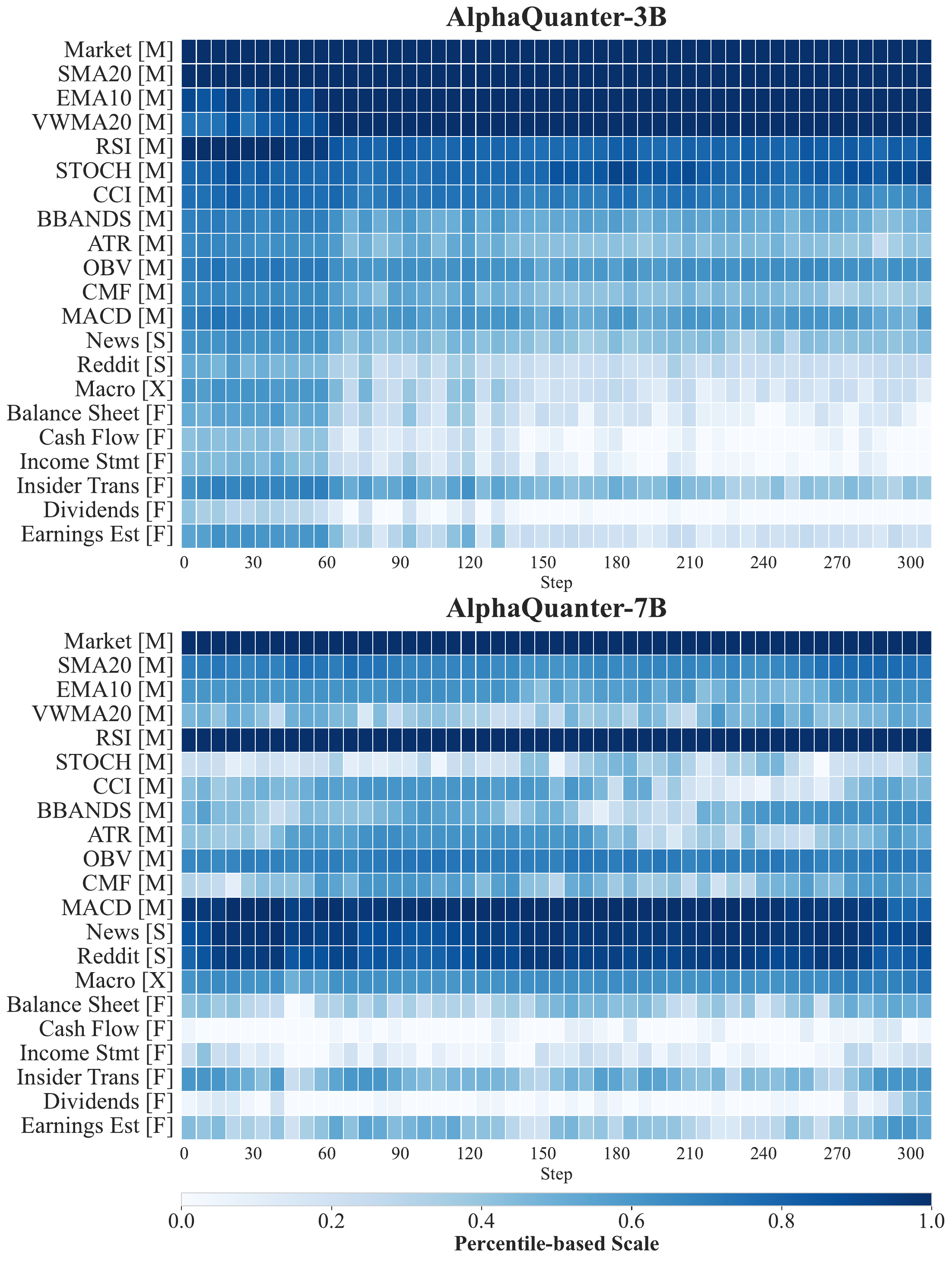}
    \vspace{-10pt}
    \caption{Evolution of the tool selection strategies for the AlphaQuanter-3B and -7B models during training. The heatmap color intensity shows the percentile-based reliance on each information at different training steps. The symbols [M], [S], [X], and [F] represent the four categories of data sources, respectively.}
    \label{fig:usage_heatmap}
    \vspace{-10pt}
\end{figure}

\begin{figure}[t]
    \centering
    \includegraphics[width=0.95\linewidth]{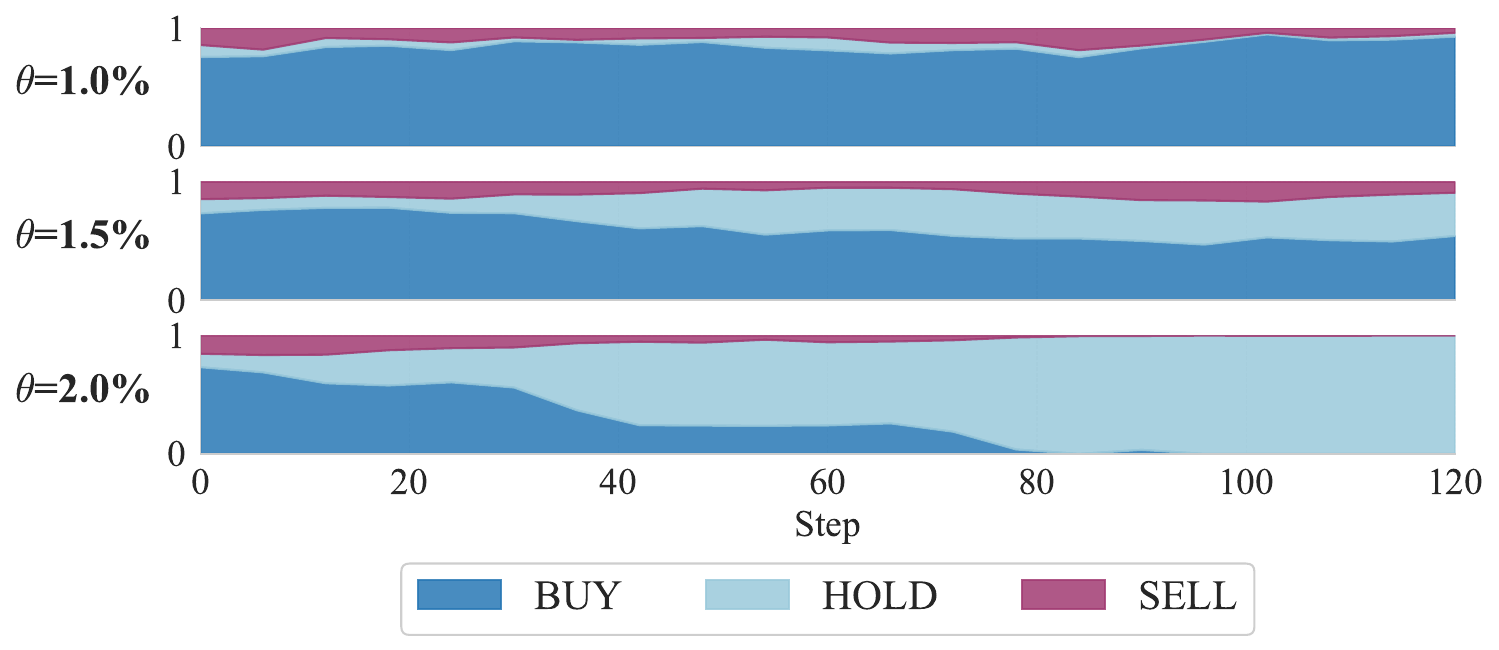}
    \vspace{-10pt}
    \caption{The effect of different decision threshold ($\theta$) values on the agent's action distribution during training.}
    \label{fig:aggregated_ratios}
    \vspace{-10pt}
\end{figure}

\paragraph{Policy Evolution}
To better understand how AlphaQuanter achieves its performance, we examine the agent's decision traces and tool-use patterns. The heatmaps in Figure~\ref{fig:usage_heatmap} show that tool usage for both the 3B and 7B models is dynamic rather than static, evolving over the course of training. This confirms that the agents actively learn and refine their information-seeking policies instead of relying on a fixed routine.

\paragraph{Divergent Strategies: 3B vs. 7B}
The two models exhibit divergent learned strategies. 
The 3B model exhibits a diffuse, low-contrast usage pattern across tools, suggesting limited ability to distinguish informative from uninformative signals. In contrast, the 7B model develops a concentrated, high-contrast pattern, consistent with a selective, discriminative policy for prioritizing information.

\begin{table}[t]
    \centering
    \scalebox{0.6}{
    \begin{tabular}{p{3.2cm}P{3.4cm}P{1.9cm}P{2.2cm}}
        \toprule
        \textbf{Model}  & {\textbf{ARR} ($\uparrow$)}  & {\textbf{SR} ($\uparrow$)} & {\textbf{MDD} ($\downarrow$)}  \\
        \midrule
AlphaQuanter-7B & \textbf{34.94\%} & \textbf{0.65} & 24.93\% \\
\midrule
$\Diamond$ w/o $\mathcal{R}_\text{format}$ & 16.36\% ($\textcolor{dGreen}{\downarrow_{53.2\%}}$) & 0.40 & 26.49\% \\
$\Diamond$ w/o $\mathcal{R}_\text{tool}$ & 19.90\% ($\textcolor{dGreen}{\downarrow_{43.0\%}}$)& 0.49 & 24.08\% \\
\midrule
$\Diamond$ $\theta \uparrow_{\change{0.5\%}}$ & 21.25\% ($\textcolor{dGreen}{\downarrow_{39.2\%}}$) & 0.28 & \ \  \textbf{9.18\%} \\
$\Diamond$ $\theta \downarrow_{\change{0.5\%}}$  & 20.23\% ($\textcolor{dGreen}{\downarrow_{42.1\%}}$) & 0.43 & 32.67\% \\
\bottomrule
    \end{tabular}}
    \caption{Impact of reward components and the threshold $\theta$ on the performance of the AlphaQuanter-7B model.} 
    \label{tab:ablation_study}
    \vspace{-10pt}
\end{table}

\paragraph{Expert-like Heuristic}
Closer examination of the 7B model's learned policy reveals a sophisticated, expert-level heuristic. It learns to rely heavily on \textit{trend}, \textit{momentum}, and \textit{volume indicators} as primary signals, while treating \textit{sentiment} and \textit{macroeconomic context} as secondary but important inputs for decision-making. At the same time, it largely downweights low-frequency \textit{fundamental} data, likely because such signals add limited value to the rapid decisions required by the task.

\subsection{Ablation Studies}
\label{sec:ablation}
We conduct an ablation study to validate the contributions of our key designs, with all results shown in Table~\ref{tab:ablation_study}. First, we evaluate the effectiveness of the process score by selectively removing the format score $\mathcal{R}_\text{format}$ and the tool score $\mathcal{R}_\text{tool}$. Their removal causes the average ARR to drop by 53.2\% and 43.0\%, respectively, confirming their critical roles in guiding the agent toward an effective and structured decision-making process. 
Next, we evaluate the sensitivity of the decision threshold $\theta$. Perturbing $\theta$ by $\pm$0.5\% yields substantial ARR reductions of 42.1\% and 39.2\%. \change{Note that $\pm0.5\%$ equals a $\pm33\%$ relative change (w.r.t.\ $\theta=1.5\%$)}. We also observe a distinct trade-off with MDD: larger $\theta$ induces more \texttt{HOLD} signals, lowering trading frequency and MDD, whereas smaller $\theta$ increases both activity and risk. As shown in Figure~\ref{fig:aggregated_ratios}, $\theta$ is crucial for balancing exploration against exploitation; an improperly calibrated value causes the agent to converge on a single action (e.g., only \texttt{BUY} or \texttt{HOLD}), whereas our setting maintains a dynamic, adaptive policy in noisy financial environments.

\section{Conclusion}

\change{In this paper, we present \textbf{AlphaQuanter}, a single-agent framework that leverages RL to optimize the decision-making process. On five large-cap U.S. stocks, AlphaQuanter improves ARR while keeping token cost per decision low, supporting practical deployment. Unlike single-agent baselines that ingest all signals at once, AlphaQuanter learns when and what to retrieve, treating tool usage as a policy action. It also avoids the conflicting signals and debate redundancy common in multi-agent systems, producing more consistent traces with higher \textit{alignment} and \textit{conciseness}, lowering the cost of manual review and verification. 
Overall, AlphaQuanter offers an effective and practical recipe for training tool-using trading policies that balance profitability, efficiency, and trace quality.}

\section*{Limitations}
While AlphaQuanter demonstrates promising results, we acknowledge several limitations. First, the agent's capabilities are bounded by its pre-defined toolset. It can orchestrate existing functions but cannot generate novel analytical methods. Second, our framework focuses on single-asset decision making, and an important extension of this work is to broaden its scope to the portfolio level. Finally, due to computational constraints, the iterative workflow can lead to an excessively long input context as observations from multiple tool calls accumulate. This highlights the need for a sophisticated agent memory mechanism~\cite{DBLP:journals/corr/abs-2404-13501} to summarize and manage the reasoning history more efficiently.

\section*{Ethics Statement}
Developing AI trading agents like AlphaQuanter that autonomously invoke tools and reasoning carries significant ethical responsibilities.
Although this technology may ultimately provide a significant advantage to well-resourced institutions, the scope of our current research is strictly confined to a simulated backtesting environment using historical data. As a research prototype, our work does not directly introduce new risks to live markets or participants. Furthermore, a core principle of our work is to enhance transparency by moving away from opaque, black-box models, which we view as a greater ethical concern. In accordance with ethical data processing guidelines, all financial data used in this research is obtained from publicly available APIs and used in accordance with relevant terms of service and license agreements. Therefore, to the best of the authors' knowledge, we believe this work introduces no additional risk.
\bibliography{custom}

\appendix
\clearpage
\newpage
\begin{center}
    {\Large\textbf{Appendices}}
\end{center}

\section{Detailed Information Sources}
\label{appx:info_source}

\subsection{Market Data}
Market data consists of two tiers: raw price/volume, and a curated set of popular technical indicators. This selection is carefully designed to offer the agent a comprehensive and non-redundant toolkit for analyzing market dynamics. These data are extracted via API from Yahoo Finance\footnote{\url{https://developer.yahoo.com/api/}} and Alpha Vantage\footnote{\url{https://www.alphavantage.co/documentation/}}.

\subsubsection{Price and Volume Data}
This data consists of the daily Open, High, Low, and Close (OHLC) prices, the Adjusted Close price, and Volume for each stock, which represents the fundamental record of an asset's trading activity for a given day.

\subsubsection{Technical Indicators}
To facilitate a deeper analysis of the raw price data, we allow the agent to query a set of the most widely used technical indicators, which we group by their analytical function:
\paragraph{Trend Indicators} These help identify the direction and strength of a price trend.
\begin{itemize}
    \item \textbf{SMA(20)}: A 20-day Simple Moving Average of the price.
    \item \textbf{EMA(10)}: A 10-day Exponential Moving Average, giving more weight to recent prices.
    \item \textbf{VWMA(20)}: A 20-day Volume-Weighted Moving Average, emphasizing periods with higher trading volume.
\end{itemize}

\paragraph{Momentum Indicators} These measure the speed and change of price movements to identify overbought or oversold conditions.
\begin{itemize}
    \item \textbf{RSI(14)}: The 14-day Relative Strength Index.
    \item \textbf{STOCH(14, 3, 3)}: The Stochastic Oscillator with the parameters defining its calculation: 14 sets the look-back period for the high-low price range, the first 3 is the smoothing period for the main oscillator line (\%K), and the second 3 is the moving average period for its signal line (\%D).
    \item \textbf{CCI(21)}: The 21-day Commodity Channel Index.
\end{itemize}
\paragraph{Volatility Indicators} These quantify the magnitude of price fluctuations.
\begin{itemize}
    \item \textbf{BBANDS(20, 2)}: Bollinger Bands. The parameter 20 sets the period for the SMA that forms the middle band. The 2 specifies that the upper and lower bands are plotted at two standard deviations above and below this middle band, respectively.
    
    \item \textbf{ATR(14)}: The 14-day Average True Range, a measure of market volatility.
\end{itemize}
\paragraph{Volume Indicators} These use trading volume to confirm trends or signal potential reversals.
\begin{itemize}
    \item \textbf{OBV}: On Balance Volume, which relates price changes to volume.
    \item \textbf{CMF}: The Chaikin Money Flow, which measures money flow volume over a period.
\end{itemize}
\paragraph{Hybrid Indicator}
\begin{itemize}
    \item \textbf{MACD(12, 26, 9)}: The Moving Average Convergence Divergence, calculated by subtracting the 26-period EMA from the 12-period EMA. The 9 refers to a 9-period EMA of the MACD line itself, which serves as a ``signal line'' to generate trading triggers.
\end{itemize}

\subsection{Fundamental Data}
To ground the agent's reasoning in a company's intrinsic financial health and valuation, we integrated several categories of fundamental data extracted via Alpha Vantage API. These sources provide a holistic view, covering core financial statements, forward-looking analyst expectations, and significant corporate events. The specific data types are detailed below:

\begin{itemize}
\item \textbf{Earnings Estimates}: Forward-looking analyst projections, including annual and quarterly estimates for Earnings Per Share (EPS) and revenue. This dataset also provides metadata such as the number of contributing analysts and their revision histories.
\item \textbf{Income Statement}: Annual and quarterly income statements detailing a company's revenues, expenses, and profitability.
\item \textbf{Balance Sheet}: Annual and quarterly balance sheets providing a snapshot of a company's assets, liabilities, and shareholders' equity.
\item \textbf{Cash Flow}: Annual and quarterly cash flow statements that report the flow of cash from operating, investing, and financing activities, normalized to standard accounting principles.
\item \textbf{Insider Transactions}: Data on historical and recent transactions of company stock executed by key stakeholders, such as executives and board members, which can serve as a signal of internal sentiment.
\item \textbf{Dividends}: A record of historical dividend payments and future declared distributions, offering insight into a company's policy on returning capital to shareholders.
\end{itemize}

\subsection{Sentiment Data}
We incorporated two sources of sentiment data: news and Reddit. 
\begin{itemize}
    \item For news data, we source headlines, summaries, and associated sentiment scores for each stock from the Alpha Vantage API.
    \item For Reddit data, we retrieve the most relevant submissions from a publicly available Reddit data dump\footnote{\url{https://academictorrents.com/details/ba051999301b109eab37d16f027b3f49ade2de13}}, focusing on content from the 11 most popular stock-trading subreddits (e.g., \texttt{wallstreetbets}, \texttt{stocks}, \texttt{Daytrading}, etc.). To manage the input context length, the content of each original post was then summarized using the Qwen3-30B-A3B-Instruct model~\cite{DBLP:journals/corr/abs-2505-09388}.
\end{itemize}

\subsection{Macroeconomic Indicators}
We integrate a set of key macroeconomic indicators extracted from the Alpha Vantage API. These indicators provide context on monetary policy, inflation, and the health of the real economy. The specific data sources are as follows:
\begin{itemize}
\item \textbf{Treasury Yield}: Data on the yields of U.S. Treasury securities across maturities, considered as a benchmark for risk-free interest rates and future economic growth expectations.
\item \textbf{Federal Funds Rate}: The target interest rate set by the U.S. Federal Reserve, provided every month. This is a primary driver of monetary policy and affects borrowing costs throughout the economy.
\item \textbf{Consumer Price Index (CPI)}: Monthly data reflecting the average change in prices paid by consumers for a basket of goods and services, serving as a primary measure of inflation.
\item \textbf{WTI Crude Oil Price}: The spot price of West Texas Intermediate crude oil. It reflects global energy prices, supply-demand dynamics, and inflationary pressures.
\item \textbf{Copper Price}: The spot price of copper, a critical industrial metal often considered a leading indicator of global economic health and manufacturing activity.
\end{itemize}

\section{Implementation Details} 
\label{app:training_details}

\subsection{Prompt Design}
\label{app:prompt_design}
We design the prompt to promote flexible, evidence-driven exploration rather than predetermined outputs~\cite{DBLP:conf/emnlp/DengC00FZYS24,DBLP:conf/emnlp/ChanJYDF0L0WS24}. This design is also consistent with recent perspectives that emphasize the role of structured intermediate representations in agentic AI systems~\cite{DBLP:journals/corr/abs-2508-12257}, as well as the broader shift from automation-oriented pipelines toward increasingly autonomous, tool-mediated reasoning workflows~\cite{DBLP:conf/emnlp/ZhengDTWBWS25}. To achieve this, we first provide a clear task description, the specific target stock and date, and available tools, with constraints on the maximum number of tool calls to ensure efficiency. Then, we enforce the designed workflow by instructing the agent to form and test hypotheses, call only one tool at a time, and clearly show its thinking process within a structured format before each action. The complete prompt is provided in the Figure~\ref{fig:full_example}.

\begin{figure*}[!h]
\begin{tcolorbox}[colback = cBlue_1!5, colframe = cBlue_6,  coltitle=white,fonttitle=\bfseries\small,fontupper=\small,fontlower=\scriptsize]

      \texttt{You are a professional trading strategy analyst. Your goal is to generate a well-reasoned final trade decision (BUY/SELL/HOLD) for a given stock and date through systematic, evidence-based exploration using all available tools. At most 8 tool calls.\\ \\
You have access to the following tools -- use them intentionally and iteratively to test hypotheses and deepen your analysis: \\ \\
- [MUST] get\_market\_data (historical OHLCV)\\
- [MUST] get\_stock\_indicators (trend indicators(SMA20, EMA10, VWMA20), momentum (RSI, \\STOCH, CCI), volatility (BBANDS, ATR), and volume-based (OBV, CMF), and hybrid(MACD))\\
- [OPTIONAL] get\_news\_data\\
- [OPTIONAL] get\_reddit\_data\\
- [OPTIONAL] get\_macro\_indicators\\
- [OPTIONAL] get\_balance\_sheet\\
- [OPTIONAL] get\_cashflow\\
- [OPTIONAL] get\_income\_statements\\
- [OPTIONAL] get\_insider\_transactions\\
- [OPTIONAL] get\_dividends\\
- [OPTIONAL] get\_earnings\_estimate \\ \\
GUIDELINES:\\
\#\# Think Like an Analyst, Not a Script.\\
Approach the problem creatively. There is no single fixed workflow. Use your reasoning to form hypotheses, then leverage tools flexibly to explore, validate, or refute your ideas. Be curious and iterative. \\ \\
\#\# Start with a High-Level Hypothesis.\\
Begin by outlining your initial perspective and what you aim to investigate. This isn't a rigid plan-it's a starting point. You're encouraged to adapt as new evidence emerges.\\ \\
\#\# Plan, Execute, Then Analyze in the format: <think> ... </think>\\
- First, Briefly Plan: Before calling any tool, briefly state your current hypothesis or what you aim to learn with the next step.\\
- Then, Call One Tool: Execute only one tool call per step. You must wait for and receive the result before proceeding.\\
- Finally, Analyze and Adapt: Interpret the result. Does it confirm your hypothesis? Does it reveal something new? Use this insight to refine your next step.\\ \\
\#\# One Step at a Time.\\
You are strictly permitted to make only one tool call at a time. The subsequent analysis and planning must be based on the returned result before any further tool is called. This ensures a deliberate and evidence-driven investigative process.\\ \\
\#\# Conclude with a Decision.\\
After synthesizing all evidence, provide a clear and justified trade recommendation in the format: <answer>BUY | SELL | HOLD</answer>\\ \\
- Current date: \{date\}\\
- Target stock ticker: \{stock\}
}

\end{tcolorbox}
    \caption{Full prompt for the AlphaQuanter agent.}
    \vspace{-5pt}
    \label{fig:full_example}
    \vspace{-5pt}
\end{figure*}

\subsection{\change{Process Score Computation}}
\label{app:process_score_computation}
\change{We provide the detailed pseudocode for computing the process score used in our reward function. Specifically, the process score is decomposed into two components: a FormatScore that evaluates whether the model follows the required response structure (e.g., valid action label, presence of reasoning tags, and appropriate token-length range), and a ToolScore that evaluates the quality of tool usage (e.g., valid arguments, duplicate calls, and reasonable tool-call counts). Please refer to Algorithm~\ref{alg:format-score}~(FormatScore) and Algorithm~\ref{alg:tool-score}~(ToolScore) for the exact scoring procedures, which are consistent with our released implementation and are used throughout RL training.}

\begin{algorithm*}[t]
\small
\caption{\textsc{FormatScore}$(\texttt{action}, \texttt{solution\_str})$}
\label{alg:format-score}
\DontPrintSemicolon
\SetKwProg{Fn}{Function}{:}{}
\SetKwFunction{ScoreResponse}{ScoreResponse}
\SetKwFunction{CountTokens}{CountTokens}

\BlankLine
\texttt{format\_score} $\leftarrow 0$\;

\BlankLine
\If(\tcp*[f]{Validate the action label}){\texttt{action} $\notin \{\texttt{"buy"}, \texttt{"sell"}, \texttt{"hold"}\}$}{
    \texttt{format\_score} $\leftarrow$ \texttt{format\_score} $- 1.0$\;
}
    \tcp{(1) Content before the first \texttt{<tool\_call>}}
    \If{\texttt{text} contains a match before the first \texttt{<tool\_call>}}{
        \texttt{segment} $\leftarrow$ \texttt{before\_tool\_call\_text}\;
        \texttt{len} $\leftarrow$ \CountTokens(\texttt{segment})\;

        \If{\texttt{text} contains both \texttt{<think>} and \texttt{</think>}}{
            \texttt{score} $\leftarrow$ \texttt{score} $+ 0.005$\;
        }

        \If{$200 \le \texttt{len} \le 600$}{
            \texttt{score} $\leftarrow$ \texttt{score} $+ 0.1$\;
        }
        \Else{
            \texttt{score} $\leftarrow$ \texttt{score} $+ 0.05 - 0.01 \cdot {(\texttt{len}-200)(\texttt{len}-600)}/{30000}$\;
        }
    }
    \Else{
        \texttt{score} $\leftarrow$ \texttt{score} $+ 0.001$\;
    }

    \BlankLine
    \tcp{(2) Content between each \texttt{\textbackslash nassistant\textbackslash n} and the next \texttt{<tool\_call>}}
    \ForEach{\texttt{match} in regex \texttt{\textbackslash nassistant\textbackslash n(.*?)<tool\_call>}}{
        \If{\texttt{match} is not empty}{
            \texttt{len} $\leftarrow$ \CountTokens(\texttt{match})\;

            \If{\texttt{match} contains both \texttt{<think>} and \texttt{</think>}}{
                \texttt{score} $\leftarrow$ \texttt{score} $+ 0.005$\;
            }

            \If{$200 \le \texttt{len} \le 600$}{
                \texttt{score} $\leftarrow$ \texttt{score} $+ 0.1$\;
            }
            \Else{
                \texttt{score} $\leftarrow$ \texttt{score} $+ 0.05 - 0.01 \cdot {(\texttt{len}-200)(\texttt{len}-600)}/{30000}$\;
            }
        }
        \Else{
            \texttt{score} $\leftarrow$ \texttt{score} $+ 0.001$\;
        }
    }

\Return \texttt{format\_score}\;
\end{algorithm*}

\begin{algorithm*}
\small
\caption{\textsc{ToolScore}$(\texttt{solution\_str}, \texttt{extra\_info}, \texttt{expected\_tool\_calls})$}
\label{alg:tool-score}
\DontPrintSemicolon
\SetKwFunction{ExtractCalls}{extract\_tool\_calls\_simple}
\SetKwFunction{CountDup}{count\_duplicate\_calls}
\SetKwFunction{CountCalls}{count\_calls}

\BlankLine
\texttt{tool\_score} $\leftarrow 0$\;
     \tcp{(1) Extract tool calls and apply initial penalty}
    $(\texttt{calls}, \texttt{initial\_penalty}) \leftarrow$ 
    \ExtractCalls(\texttt{solution\_str})\;
    \texttt{tool\_score} $\leftarrow$ \texttt{tool\_score} $+$ \texttt{initial\_penalty}\;
    \tcp{Note: \ExtractCalls{} applies, for each JSON-list block: }
    \tcp{$\qquad \texttt{penalty} \mathrel{-}= |\texttt{tool\_list}|^2/10$.}
    \tcp{(2) Validate date arguments in each call}
    \ForEach{\texttt{call} in \texttt{calls}}{
        \texttt{arg\_date} $\leftarrow$ \texttt{call["arguments"]["curr\_date"]}\;

        \If{\texttt{arg\_date} $>$ \texttt{extra\_info["date"]}}{
            \texttt{result\_reward} $\leftarrow -1$\;
            \texttt{tool\_score} $\leftarrow$ \texttt{tool\_score} $- 50.0$\;
        }
    }
    \tcp{(3) Penalize duplicate tool calls}
    \texttt{duplicate\_count} $\leftarrow$ \CountDup(\texttt{calls})\;
    \texttt{tool\_score} $\leftarrow$ \texttt{tool\_score} $- 0.05 \cdot \texttt{duplicate\_count}$\;
    \tcp{(4) Penalize tool call count outside the preferred range}
    \texttt{num\_calls} $\leftarrow$ \CountCalls(\texttt{calls})\;
    \If{\texttt{num\_calls} $\le 4$ \textbf{or} \texttt{num\_calls} $> 8$}{
        \texttt{tool\_score} $\leftarrow$ \texttt{tool\_score} $- \left|\texttt{num\_calls} - \texttt{expected\_tool\_calls}\right|$\;
    }
\Return \texttt{tool\_score}\;
\end{algorithm*}

\subsection{Hyperparameters for RL Training}
We train AlphaQuanter using verl~\cite{DBLP:conf/eurosys/ShengZYWZZPL025}. In Table~\ref{tab:hpp}, we list the important parameter settings for the verl framework as well as the hyperparameters referenced in this paper.
\begin{table}[h]
    \centering
    \renewcommand{\arraystretch}{1.1}
    \scalebox{0.6}{
    \begin{tabular}{L{11cm}L{1cm}}
        \toprule
        \textbf{Key}  & \textbf{Value} \\
        \midrule
        \texttt{algorithm.use\_kl\_in\_reward} & false \\
\texttt{actor\_rollout\_ref.actor.clip\_ratio\_low} & 0.1 \\
\texttt{actor\_rollout\_ref.actor.clip\_ratio\_high} & 0.1 \\
\texttt{actor\_rollout\_ref.actor.clip\_ratio\_c} & 3 \\
\texttt{actor\_rollout\_ref.actor.entropy\_coeff} & 0 \\
\texttt{actor\_rollout\_ref.actor.kl\_loss\_coef} & 0.05 \\
\texttt{actor\_rollout\_ref.actor.optim.lr} & 1e-6 \\
\texttt{actor\_rollout\_ref.actor.use\_kl\_loss} & true \\
\texttt{actor\_rollout\_ref.rollout.multi\_turn.max\_user\_turns} & 32 \\
\texttt{actor\_rollout\_ref.rollout.multi\_turn.max\_assistant\_turns} & 32 \\
\texttt{actor\_rollout\_ref.rollout.n} & 16 \\
\texttt{algorithm.kl\_ctrl.kl\_coef} & 0.0 \\
\texttt{data.max\_prompt\_length} & 3072 \\
\texttt{data.max\_response\_length} & 16384 \\
\texttt{data.train\_batch\_size} & 32 \\
\midrule
$H$  & 7\\
$\lambda$ & 0.001\\
$\kappa$ & 0.9 \\
$\theta$  & 0.015\\
$\alpha$  & 5\\
$\min_\text{token}$ & 200 \\
$\max_\text{token}$ & 600 \\
$\min_\text{tool}$ & 4 \\
$\max_\text{tool}$ & 8 \\
\bottomrule
    \end{tabular}}
    \caption{Hyperparameters for training AlphaQuanter.} 
    \label{tab:hpp}
    \vspace{-15pt}
\end{table}

\begin{table*}[!t]
    \centering
    \renewcommand{\arraystretch}{1.08}
    \scalebox{0.68}{
    \begin{tabular}{P{3.2cm}L{3.2cm}
                    R{1.5cm}R{1.10cm}R{1.5cm}
                    R{1.5cm}R{1.10cm}R{1.5cm}
                    R{1.5cm}R{1.10cm}R{1.5cm}}
        \toprule
        \multirow{2}{*}{\textbf{Category}} & \multirow{2}{*}{\textbf{Model}}
        & \multicolumn{3}{c}{\textbf{GOOGL}}
        & \multicolumn{3}{c}{\textbf{META}}
        & \multicolumn{3}{c}{\textbf{MSFT}} \\
        \cmidrule(r){3-5}  \cmidrule(r){6-8} \cmidrule(r){9-11}
        & & \textit{ARR} ($\uparrow$) & \textit{SR} ($\uparrow$) & \textit{MDD} ($\downarrow$)
          & \textit{ARR} ($\uparrow$) & \textit{SR} ($\uparrow$) & \textit{MDD} ($\downarrow$)
          & \textit{ARR} ($\uparrow$) & \textit{SR} ($\uparrow$) & \textit{MDD} ($\downarrow$) \\
        \midrule

        \textbf{Market} & B\&H
        & -14.49\% & -0.35 & 27.35\%
        & 45.64\%  & 1.25  & 31.59\%
        & 36.80\%  & 1.41  & 18.79\% \\
        \midrule

        \multirow{2}{*}{\textbf{Rule}} & MACD
        & -3.17\% & -0.04 & 14.14\%
        & 46.82\% & 2.17  & 12.51\%
        & -9.58\% & -0.49 & 19.97\% \\
        & ZMR
        & -2.26\% & 0.01 & 18.47\%
        & -0.98\% & 0.12 & 15.19\%
        & 8.53\%  & 0.56 & 9.59\% \\
        \midrule

        \multirow{2}{*}{\textbf{RL}} & FinRL\textsubscript{A2C}
        & -21.22\% & -0.50 & 29.80\%
        & 43.41\%  & 1.04  & 34.14\%
        & 43.14\%  & 1.35  & 20.55\% \\
        & FinRL\textsubscript{PPO}
        & -19.77\% & -0.43 & 29.80\%
        & 50.34\%  & 1.15  & 34.15\%
        & \textcolor{dOra}{\textbf{43.90\%}}  & 1.36 & 20.55\% \\
        \midrule

\multirow{1}{*}{{\textbf{LM}}} & Chronos-2
        & \textcolor{dRed}{\textbf{19.07\%}}   & 0.60   & 19.43\%
        & -12.61\%  & -0.21  & 36.59\%
        & 20.04\%   & 0.64   & 13.73\% \\

\midrule
        \multirow{2}{*}{\textbf{LLM}} & FinMem
        & -22.41\% & -0.38 & 29.80\%
        & 46.25\%  & 0.77  & 34.14\%
        & 40.26\%  & 0.90  & 20.55\% \\
        & TradingAgents
        & -14.95\% & -0.29 & 25.93\%
        & 29.69\%  & 0.71  & 14.05\%
        & 38.62\%  & 0.90  & 19.83\% \\
        \midrule

        \multirow{7}{*}{\makecell{\textbf{Multi-Agent}\\ \footnotesize({TradingAgents})}}
        & Qwen2.5-3B
        & 1.73\%  & 0.10 & 5.52\%
        & 36.25\% & 0.85 & 15.28\%
        & 40.89\% & 1.06 & 12.23\% \\
        & Qwen2.5-7B
        & \textcolor{dOra}{\textbf{9.33\%}} & 1.38 & 1.40\%
        & 28.98\% & 0.87 & 6.54\%
        & -4.50\% & -1.05 & 2.27\% \\
        & Qwen3-30B-A3B
        & -18.09\% & -0.46 & 26.36\%
        & 1.36\%   & 0.29  & 16.29\%
        & 9.84\%   & 0.42  & 15.88\% \\
        & DeepSeek-V3.1\textsubscript{\textit{685B}}
        & -12.43\% & -0.66 & 12.01\%
        & -9.48\%  & -0.25 & 17.18\%
        & 14.13\%  & 0.60  & 10.09\% \\
        & Kimi-K2\textsubscript{\textit{1T}}
        & -23.40\% & -1.09 & 17.57\%
        & -9.52\%  & -0.10 & 16.12\%
        & 12.60\%  & 0.51  & 9.11\% \\
        & GPT-4o-mini
        & -18.08\% & -0.94 & 18.86\%
        & 0.73\%   & 0.04  & 11.11\%
        & 16.27\%  & 0.48  & 18.52\% \\
        & GPT-4o
        & -14.95\% & -0.29 & 25.93\%
        & 29.69\%  & 0.71  & 14.05\%
        & 38.62\%  & 0.90  & 19.83\% \\
        \midrule

        \multirow{7}{*}{\makecell{\textbf{Single-Agent}\\ \footnotesize({AlphaQuanter w/o RL})}}
        & Qwen2.5-3B
        & {{3.06\%}} & 0.07 & 18.18\%
        & 23.08\% & 0.52 & 24.91\%
        & 5.10\%  & 0.14 & 14.66\% \\
        & Qwen2.5-7B
        & -22.42\% & -0.43 & 28.59\%
        & 35.50\%  & 0.56  & 28.49\%
        & 17.55\%  & 0.48  & 19.60\% \\
        & Qwen3-30B-A3B
        & -26.33\% & -0.50 & 28.39\%
        & 32.86\%  & 0.81  & 28.18\%
        & 37.45\%  & 0.87  & 21.15\% \\
        & DeepSeek-V3.1\textsubscript{\textit{685B}}
        & -25.15\% & -0.47 & 29.77\%
        & 32.49\%  & 0.61  & 34.14\%
        & 25.45\%  & 0.64  & 19.94\% \\
        & Kimi-K2\textsubscript{\textit{1T}}
        & -40.48\% & -0.39 & 24.67\%
        & 25.83\%  & 0.68  & 21.65\%
        & -3.39\%  & -0.03 & 19.21\% \\
        & GPT-4o-mini
        & -24.02\% & -0.56 & 23.20\%
        & 44.42\%  & 0.97  & 23.84\%
        & 43.42\%  & 1.10  & 12.92\% \\
        & GPT-4o
        & -9.01\% & -0.12 & 19.72\%
        & \textcolor{dRed}{\textbf{57.18\%}} & 0.99 & 25.02\%
        & 19.39\% & 0.53 & 23.04\% \\
        \midrule

        \multirow{2}{*}{\makecell{\textbf{Single-Agent + RL}\\ \footnotesize({AlphaQuanter})}}
        & AlphaQuanter-3B
        & -14.68\% & -0.29 & 25.60\%
        & \textcolor{dOra}{\textbf{56.15\%}}  & 1.08  & 23.75\%
        & 9.82\%   & 0.30  & 21.06\% \\
        & AlphaQuanter-7B
        & -2.52\% & 0.05 & 21.37\%
        & 41.91\% & 0.78 & 25.65\%
        & \textcolor{dRed}{\textbf{47.23\%}} & 1.17 & 14.85\% \\
        \bottomrule
    \end{tabular}}
    \caption{Performance comparison of different methods over a six-month backtesting period (1/2): detailed results for [GOOGL, META, MSFT].}
    \label{tab:app:result_122_1}
\end{table*}

\subsection{Detail of Baseline}
\change{For the \textbf{Market} baseline, we implement a passive buy-and-hold strategy by having the agent output \texttt{BUY} on every trading day. For \textbf{Rule} baseline, we implement two standard technical strategies. The first is Moving Average Convergence Divergence (MACD), a trend-following strategy that uses indicator crossovers to generate trading signals; we employ the standard (12, 26, 9) parameterization for the fast, slow, and signal periods. The second is Z-score Mean Reversion (ZMR), which assumes price reversion to a historical mean. We enter a trade when the Z-score (calculated over a 20-period lookback) exceeds a threshold of 1.0 and exits upon reversion to the mean (Z-score = 0). For \textbf{RL} baselines, we include two widely used deep RL trading agents, A2C and PPO, implemented in the FinRL framework~\cite{DBLP:journals/corr/abs-2011-09607}  and trained with default hyperparameters on the same five stocks under an identical backtesting protocol.
For \textbf{LM} baseline, we implement Chronos-2~\cite{DBLP:journals/corr/abs-2510-15821} by feeding the past 30 days of adjusted closing prices to predict the next-day price, and the predicted return is then used to derive trading decisions.
For \textbf{LLM} baselines, we compare against representative prompt-based LLM traders, including FinMem~\cite{DBLP:journals/tbd/YuLCJLSZK25} and TradingAgents~\cite{xiao2024tradingagents}. For a fair comparison, we use GPT-4o as the backbone LLM for these baselines and run them with our unified data sources/tools. For the \textbf{Multi-Agent} baseline, we adapt the framework from ~\citet{xiao2024tradingagents} and replace the original data sources with our four designated categories of financial data, while retaining the prompts and agent architecture as specified in the original paper. For the \textbf{Single-Agent} baseline, we design a configuration that utilizes our custom prompt structure, as shown in Figure~\ref{fig:full_example}, in conjunction with the same four data categories. This baseline serves to isolate the performance of a single agent with full informational access but without the RL-optimized workflow of AlphaQuanter.}

\section{Detailed Result Analysis}

\begin{table*}[!t]
    \centering
    \renewcommand{\arraystretch}{1.08}
    \scalebox{0.68}{
    \begin{tabular}{P{3.2cm}L{3.2cm}
                    R{1.5cm}R{1.10cm}R{1.5cm}
                    R{1.5cm}R{1.10cm}R{1.5cm}
                    >{\columncolor{dBlue!20}}R{1.5cm}>{\columncolor{dBlue!20}}R{1.10cm}>{\columncolor{dBlue!20}}R{1.5cm}}
        \toprule
        \multirow{2}{*}{\textbf{Category}} & \multirow{2}{*}{\textbf{Model}}
        & \multicolumn{3}{c}{\textbf{NVDA}}
        & \multicolumn{3}{c}{\textbf{TSLA}}
        & \multicolumn{3}{>{\columncolor{dBlue!20}}c}{\textbf{Average}} \\
        \cmidrule(r){3-5}  \cmidrule(r){6-8} \cmidrule(r){9-11}
        & & \textit{ARR} ($\uparrow$) & \textit{SR} ($\uparrow$) & \textit{MDD} ($\downarrow$)
          & \textit{ARR} ($\uparrow$) & \textit{SR} ($\uparrow$) & \textit{MDD} ($\downarrow$)
          & \textit{ARR} ($\uparrow$) & \textit{SR} ($\uparrow$) & \textit{MDD} ($\downarrow$) \\
        \midrule

        \textbf{Market} & B\&H
        & 25.47\% & 0.74 & 33.83\%
        & -28.91\% & -0.20 & 44.10\%
        & 12.90\% & 0.57 & 31.13\% \\
        \midrule

        \multirow{2}{*}{\textbf{Rule}} & MACD
        & -12.89\% & -0.22 & 30.76\%
        & 22.77\%  & 0.78  & 28.83\%
        & 8.79\%   & 0.44  & 21.24\% \\
        & ZMR
        & 35.01\% & 1.03 & 16.72\%
        & 16.74\% & 0.59 & 44.33\%
        & 11.41\% & 0.46 & 20.86\% \\
        \midrule

        \multirow{2}{*}{\textbf{RL}} & FinRL\textsubscript{A2C}
        & 37.42\% & 0.84 & 32.68\%
        & -35.10\% & -0.19 & 48.18\%
        & 13.53\% & 0.51 & 33.07\% \\
        & FinRL\textsubscript{PPO}
        & 18.56\% & 0.58 & 35.93\%
        & -31.94\% & -0.12 & 48.18\%
        & 12.22\% & 0.51 & 33.72\% \\
        \midrule

\multirow{1}{*}{{\textbf{LM}}} & Chronos-2
        & \textcolor{dOra}{\textbf{38.19\%}}   & 0.73   & 16.32\%
        & -17.66\%  & -0.04  & 35.62\%
        & 9.41\%   & 0.34   & 24.34\% \\

\midrule

        \multirow{2}{*}{\textbf{LLM}} & FinMem
        & 26.71\% & 0.48 & 36.89\%
        & -22.28\% & -0.27 & 24.32\%
        & 13.71\% & 0.30 & 29.14\% \\
        & TradingAgents
        & -7.83\% & 0.03 & 38.74\%
        & 36.92\% & 1.17 & 10.56\%
        & 16.49\% & 0.50 & 21.82\% \\
        \midrule

        \multirow{7}{*}{\makecell{\textbf{Multi-Agent}\\ \footnotesize({TradingAgents})}}
        & Qwen2.5-3B
        & -3.28\% & -0.06 & 18.77\%
        & -76.98\% & -2.60 & 52.95\%
        & -0.28\% & -0.13 & 20.95\% \\
        & Qwen2.5-7B
        & -17.22\% & -0.99 & 14.12\%
        & -9.11\%  & -0.59 & 7.82\%
        & 1.50\%   & -0.08 & 6.43\% \\
        & Qwen3-30B-A3B
        & 10.22\% & 0.31 & 23.78\%
        & -16.51\% & -0.25 & 28.71\%
        & -2.64\% & 0.06 & 22.20\% \\
        & DeepSeek-V3.1\textsubscript{\textit{685B}}
        & -24.02\% & -0.97 & 23.18\%
        & 0.00\%   & 0.00  & 0.00\%
        & -6.36\% & -0.26 & 12.49\% \\
        & Kimi-K2\textsubscript{\textit{1T}}
        & -8.33\% & -0.28 & 18.88\%
        & 8.88\%  & 0.40  & 71.40\%
        & -3.95\% & -0.11 & 26.62\% \\
        & GPT-4o-mini
        & -5.38\% & 0.01 & 36.61\%
        & 5.20\%  & 0.10 & 6.30\%
        & -0.25\% & -0.06 & 18.28\% \\
        & GPT-4o
        & -7.83\% & 0.03 & 38.74\%
        & \textcolor{dOra}{\textbf{36.92\%}} & 1.17 & 10.56\%
        & 16.49\% & 0.50 & 21.82\% \\
        \midrule

        \multirow{7}{*}{\makecell{\textbf{Single-Agent}\\ \footnotesize({AlphaQuanter w/o RL})}}
        & Qwen2.5-3B
        & -7.43\% & 0.14 & 34.63\%
        & -32.21\% & -0.46 & 37.59\%
        & -1.68\% & 0.08 & 25.99\% \\
        & Qwen2.5-7B
        & 1.47\% & 0.22 & 40.24\%
        & -9.63\% & -0.04 & 27.88\%
        & 4.49\% & 0.16 & 28.96\% \\
        & Qwen3-30B-A3B
        & 29.61\% & 0.51 & 33.48\%
        & -46.41\% & -1.08 & 39.22\%
        & 5.44\% & 0.12 & 30.08\% \\
        & DeepSeek-V3.1\textsubscript{\textit{685B}}
        & 10.30\% & 0.31 & 39.81\%
        & -1.21\% & 0.13 & 29.82\%
        & 8.38\% & 0.24 & 30.70\% \\
        & Kimi-K2\textsubscript{\textit{1T}}
        & -3.27\% & 0.11 & 34.92\%
        & 13.05\% & 0.36 & 26.05\%
        & -1.65\% & 0.15 & 25.30\% \\
        & GPT-4o-mini
        & 13.61\% & 0.35 & 37.60\%
        & -43.71\% & -0.59 & 36.32\%
        & 6.74\% & 0.25 & 26.78\% \\
        & GPT-4o
        & 17.60\% & 0.39 & 38.53\%
        & -38.04\% & -0.54 & 35.06\%
        & 9.42\% & 0.25 & 28.27\% \\
        \midrule

        \multirow{2}{*}{\makecell{\textbf{Single-Agent + RL}\\ \footnotesize({AlphaQuanter})}}
        & AlphaQuanter-3B
        & {{30.55\%}} & 0.51 & 29.04\%
        & 33.33\% & 0.57 & 26.34\%
        & \textcolor{dOra}{\textbf{23.03\%}} & 0.43 & 25.16\% \\
        & AlphaQuanter-7B
        & \textcolor{dRed}{\textbf{45.41\%}} & 0.66 & 34.91\%
        & \textcolor{dRed}{\textbf{42.67\%}} & 0.58 & 27.88\%
        & \textcolor{dRed}{\textbf{34.94\%}} & 0.65 & 24.93\% \\
        \bottomrule
    \end{tabular}}
    \caption{Performance comparison of different methods over a six-month backtesting period (2/2): detailed results for [NVDA, TSLA] and average.}
    \label{tab:app:result_122_2}
\end{table*}

\subsection{Full Results of Main Table}
Table~\ref{tab:app:result_122_1} and Table~\ref{tab:app:result_122_2} present the complete backtesting results, providing a detailed breakdown of the ARR, SR, and MDD for each individual stock summarized in Table~\ref{tab:main_result_122}. Our asset-specific analysis reveals several key findings. 
For {GOOGL}, most models struggle to generate positive returns, although a few baseline methods achieve marginal gains. Notably, Chronos-2 stands out on this stock by achieving the highest ARR, while its performance on the other stocks is less competitive overall. 
For {META}, the majority of strategies are profitable. Notably, the single-agent version of GPT-4o achieved the highest ARR, a result matched by AlphaQuanter-3B, which does so with a superior risk profile, evidenced by a higher SR and a lower MDD. 
On MSFT, AlphaQuanter-7B delivers the highest ARR, concurrently achieving a strong SR and a relatively low MDD. 
For NVDA, the results are mixed, with returns split between positive and negative. We observe that multi-agent methods are more prone to negative returns, whereas single-agent approaches more frequently yield positive returns with SRs greater than zero, although with high MDD. Here, AlphaQuanter-7B again secures the highest ARR, with its SR and MDD being comparable to the market baseline.
For TSLA, the performance is similarly divided. It is particularly noteworthy that DeepSeek-V3.1 consistently outputs a HOLD signal, resulting in zero values for all metrics. This behavior empirically validates our earlier assertion that prompting-based models struggle to differentiate between BUY and HOLD signals under uncertainty. AlphaQuanter-7B once again achieves the highest ARR with satisfactory SR and MDD.

\subsection{Rolling-Window Robustness Analysis}
\begin{table*}[h]
    \centering
    \renewcommand{\arraystretch}{1.1}
    \scalebox{0.7}{
        \begin{tabular}{L{4.3cm}R{1.8cm}R{1.8cm}R{1.8cm}R{1.8cm}R{1.8cm}>{\columncolor{dBlue!20}}R{1.8cm}>{\columncolor{dBlue!20}}R{1.8cm}>{\columncolor{dBlue!20}}R{1.8cm}}
        \toprule
         \multirow{2}{*}{\textbf{Model}}  & {\textbf{GOOGL}}  & {\textbf{META}} & \textbf{MSFT} & \textbf{NVDA} & \textbf{TSLA}  & \multicolumn{3}{>{\columncolor{dBlue!20}}c}{\textbf{Average}} \\
         \cmidrule(r){2-2} \cmidrule(r){3-3} \cmidrule(r){4-4} \cmidrule(r){5-5} \cmidrule(r){6-6} \cmidrule(r){7-9}
         & \textit{ARR} ($\uparrow$) & \textit{ARR} ($\uparrow$)  & \textit{ARR} ($\uparrow$) & \textit{ARR} ($\uparrow$) & \textit{ARR} ($\uparrow$) & \textit{ARR} ($\uparrow$) & \textit{SR} ($\uparrow$) & \textit{MDD} ($\downarrow$)  \\
        \midrule
Buy \& Hold                 & -23.02\%         & 2.74\%           & 50.37\%          & 40.60\%          & 45.71\% & 23.28\%          & -0.01             & 25.01\%      \\
FinMem                         & -21.33\%         & 15.78\%          & \textbf{79.85\%} & 42.49\%          & -24.32\% & 18.49\%          & -0.08             & 21.36\%   \\
TradingAgents & -16.06\%         & 22.16\%          & 58.03\%          & -4.52\%          & 61.86\%  & 24.29\%          & 0.25              & \textbf{12.16\%}  \\
\midrule
AlphaQuanter-3B              & -27.72\%         & \textbf{46.16\%} & 22.17\%          & \textbf{78.61\%} & 88.52\%& 41.55\%          & 0.13              & 19.02\%      \\
AlphaQuanter-7B         & \textbf{9.35\%}  & 43.65\%          & 29.91\%          & 62.86\%          & \textbf{115.51\%} & \textbf{52.26\%} & \textbf{0.26}     & 17.48\%         \\

\bottomrule
    \end{tabular}}
    \caption{Rolling-window robustness evaluation on the January -- June 2025 test period. We use a 3-month rolling horizon with a 7-day step size. Results are averaged over all rolling windows.} 
    \label{tab:app:rolling_window}
\end{table*}

We further conduct a rolling-window evaluation, directly testing whether our method remains effective when the evaluation window shifts over time. Following common practice in quantitative finance, we evaluate performance over a 3-month horizon using a rolling-origin protocol~\cite{fpp3}. Specifically, on the full January--June 2025 test span, we use a 3-month rolling window and move the window forward by 7 days at each iteration. This creates multiple overlapping but diverse evaluation periods covering substantially different short-term market conditions. Importantly, all models are evaluated without any retraining or hyperparameter tuning; for AlphaQuanter, we keep the original decision threshold $\theta$ fixed across all rolling windows. Due to space constraints, we report the most competitive and representative baselines, including Buy \& Hold, FinMem, and TradingAgents.

Table~\ref{tab:app:rolling_window} summarizes the average performance across all rolling windows. AlphaQuanter-7B achieves the best overall Avg.\ ARR (52.26\%) and Avg.\ SR (0.26), while maintaining a competitive Avg.\ MDD (17.48\%). AlphaQuanter-3B also substantially improves over the representative baselines in Avg.\ ARR. At the stock level, the rolling-window results remain heterogeneous, as expected in realistic markets, but AlphaQuanter retains strong performance on multiple assets, especially META, NVDA, and TSLA. Importantly, these gains are obtained without altering $\theta$ or any other hyperparameters across windows, which reduces the likelihood that the main results are driven by a fortunate choice of the original backtesting period. Overall, the rolling-window evidence supports that AlphaQuanter learns a policy that generalizes more robustly across changing market conditions than the compared baselines.

\subsection{Reward Decomposition Analysis}
\begin{figure}[!h]
    \centering
    \includegraphics[width=\linewidth]{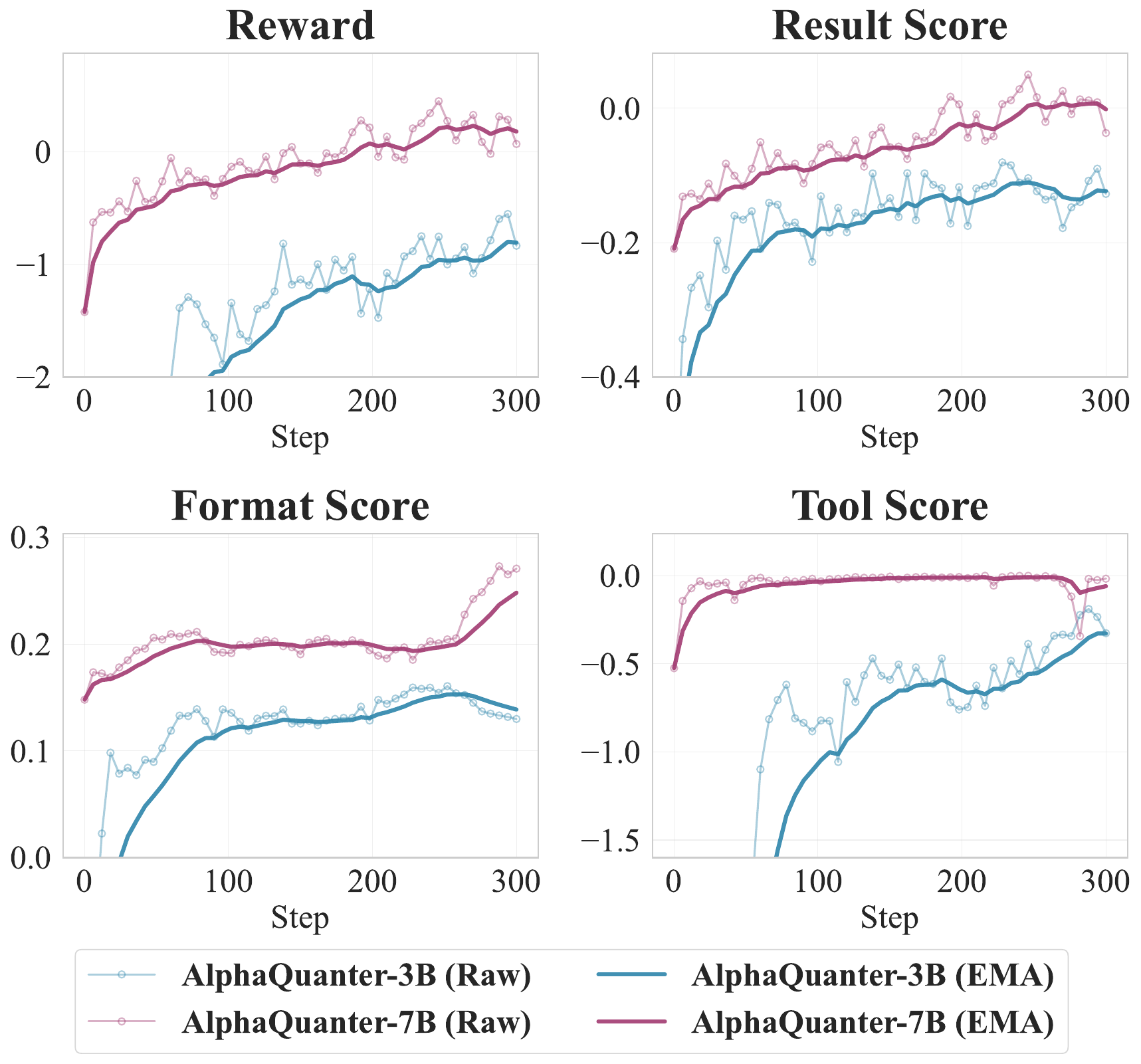}
    \caption{A comparative analysis of the training dynamics for the AlphaQuanter-3B and -7B models, illustrating the evolution of the total reward and its score components.}
    \label{fig:score_comparison}
    \vspace{-10pt}
\end{figure}

To complement the analysis in Section~\ref{sec:tdvp}, Figure~\ref{fig:score_comparison} displays the learning curves for the primary reward and its constituent components, the result, format, and tool scores, on the validation set during training. A key observation is that the 7B model consistently outperforms the 3B model across all scoring metrics. The result score exhibits a clear upward trend for both models, indicating a steady improvement in the accuracy of the agent's final actions. The rate of improvement gradually decelerates as the models converge. Regarding the format score, which reflects the length of the agent's reasoning trace, both models initially show an increase. However, after approximately 250 steps, their paths become different: the 7B model continues to generate more detailed reasoning, while the 3B model's reasoning length begins to decrease, leading to a decline in its format score. For the tool score, the 3B model initially performs poorly and incurs significant penalties. A case study of its rollouts reveals that in the early stages, the 3B model fails to adhere to instructions by making multiple tool calls within a single turn, which is the primary cause of its low score. This behavior is gradually rectified through further training.

\begin{table*}[!t]
    \centering
    \renewcommand{\arraystretch}{1.1}
    \scalebox{0.73}{
    \begin{tabular}{p{4.1cm}R{1.6cm}R{1.2cm}R{1.6cm}R{1.6cm}R{1.2cm}R{1.6cm}R{1.6cm}R{1.2cm}R{1.6cm}}
        \toprule
        \multirow{2}{*}{\textbf{Model}}  & \multicolumn{3}{c}{\textbf{GOOGL}}  & \multicolumn{3}{c}{\textbf{META}} & \multicolumn{3}{c}{\textbf{MSFT}} \\
        \cmidrule(r){2-4}  \cmidrule(r){5-7} \cmidrule(r){8-10}
        & ARR ($\uparrow$) & SR ($\uparrow$) & MDD ($\downarrow$) & ARR ($\uparrow$) & SR ($\uparrow$) & MDD ($\downarrow$) & ARR ($\uparrow$) & SR ($\uparrow$) & MDD ($\downarrow$)  \\
        \midrule
AlphaQuanter-7B & -2.52\% & 0.05 & 21.37\% & 41.91\% & 0.78 & 25.65\% & \textbf{47.23\%} & 1.17 & 14.85\% \\
\midrule
$\Diamond$ w/o $\mathcal{R}_\text{format}$ & -6.40\% & -0.09 & 24.86\% & 12.99\% & 0.66 & 25.03\% & 13.94\% & 0.51 & 18.93\% \\
$\Diamond$ w/o $\mathcal{R}_\text{tool}$ & -14.22\% & -0.25 & 25.28\% & 47.29\% & 0.85 & 24.23\% & 28.40\% & 0.72 & 19.81\% \\
\midrule
$\Diamond$ $\theta \uparrow_{0.5\%}$ & \textbf{2.83\%} & 0.10 & 4.59\% & 16.07\% & 0.27 & 10.91\% & 16.53\% & 0.48 & 2.40\% \\
$\Diamond$ $\theta \downarrow_{0.5\%}$ & -13.05\% & -0.16 & 28.66\% & \textbf{50.82\%} & 0.82 & 34.50\% & 38.16\% & 0.87 & 20.01\% \\
\bottomrule
    \end{tabular}}
    \caption{Impact of reward components and the threshold $\theta$ on the performance of the AlphaQuanter-7B model (1/2): detailed results for [GOOGL, META, MSFT].} 
    \label{tab:app:abla_1}
\end{table*}

\begin{table*}[t]
    \centering
    \renewcommand{\arraystretch}{1.1}
    \scalebox{0.73}{
    \begin{tabular}{p{4.1cm}R{1.6cm}R{1.2cm}R{1.6cm}R{1.6cm}R{1.2cm}R{1.6cm}>{\columncolor{dBlue!20}}R{1.6cm}>{\columncolor{dBlue!20}}R{1.2cm}>{\columncolor{dBlue!20}}R{1.6cm}}
        \toprule
        \multirow{2}{*}{\textbf{Model}}  & \multicolumn{3}{c}{\textbf{NVDA}}  & \multicolumn{3}{c}{\textbf{TSLA}} & \multicolumn{3}{>{\columncolor{dBlue!20}}c}{\textbf{{Average}}} \\
        \cmidrule(r){2-4}  \cmidrule(r){5-7} \cmidrule(r){8-10}
        & ARR ($\uparrow$) & SR ($\uparrow$) & MDD ($\downarrow$) & ARR ($\uparrow$) & SR ($\uparrow$) & MDD ($\downarrow$) & ARR ($\uparrow$) & SR ($\uparrow$) & MDD ($\downarrow$)  \\
        \midrule
AlphaQuanter-7B & \textbf{45.41\%} & 0.66 & 34.91\% & \textbf{42.67\%} & 0.58 & 27.88\% & \textbf{34.94\%} & 0.65 & 24.93\% \\
\midrule
$\Diamond$ w/o $\mathcal{R}_\text{format}$ & 33.70\% & 0.49 & 35.55\% & 27.59\% & 0.43 & 28.06\% & 16.36\% & 0.40 & 26.49\% \\
$\Diamond$ w/o $\mathcal{R}_\text{tool}$ & 20.73\% & 0.43 & 35.24\% & 17.28\% & 0.70 & 15.85\% & 19.90\% & 0.49 & 24.08\% \\
\midrule
$\Diamond$ $\theta \uparrow_{0.5\%}$  & 40.00\% & 0.22 & 20.88\% & 30.84\% & 0.32 & 7.14\% & 21.25\% & 0.28 & 9.18\% \\
$\Diamond$ $\theta \downarrow_{0.5\%}$  & 31.73\% & 0.53 & 36.50\% & -6.50\% & 0.11 & 43.66\% & 20.23\% & 0.43 & 32.67\% \\
\bottomrule
    \end{tabular}}
    \caption{Impact of reward components and the threshold $\theta$ on the performance of the AlphaQuanter-7B model (2/2): detailed results for [NVDA, TSLA] and average.} 
    \label{tab:app:abla_2}
\end{table*}

\subsection{\change{Detailed Faithfulness Analysis}} \label{app:detailed_faith_analysis}
\change{We provide the detailed scoring rules for the three human-rated faithfulness metrics used in Section~\ref{sec:faith_analysis}. Each output is rated independently by three Ph.D. raters, who voluntarily agree to participate in this research without receiving any compensation,
on an ordinal scale \{0, 1, 2\}, where higher is better.}
\paragraph{Reason-Tool Alignment (0-2)} \change{This metric evaluates whether the model's stated information needs in its reasoning are consistent with its executed tool calls.}

\begin{itemize}
    \item \textbf{2 (High alignment)}: The reasoning explicitly states the key information to retrieve, and the subsequent tool calls match these stated intentions.
    \item \textbf{1 (Partial alignment)}: Some key intentions are matched, but there is at least one noticeable mismatch.
    \item \textbf{0 (Low alignment)}: Frequent or severe mismatches between reasoning and tool calls, such that the tool usage appears disconnected from the stated plan.
\end{itemize}

\paragraph{Evidence Grounding (0-2)} \change{This metric evaluates whether the decision is grounded in verifiable evidence from tool outputs. For each output, the rater checks up to two key factual/diagnostic claims that are most central to the final BUY/SELL/HOLD decision. A claim is considered grounded only if it can be directly supported by the retrieved tool outputs in the same episode.}

\begin{itemize}
    \item \textbf{2 (Highly grounded)}: Both checked key claims are supported by the tool outputs.
    \item \textbf{1 (Partially grounded)}: Only one of the two checked claims is supported; the other is missing or unverifiable.
    \item \textbf{0 (Ungrounded)}: Neither checked claim is supported by the tool outputs, or the output contains central claims that contradict the retrieved evidence.
\end{itemize}

\paragraph{Conciseness (0-2)} \change{This metric evaluates whether the decision trace remains concise, avoiding redundant tool calls and verbose chatter that obscures the evidence.}

\begin{itemize}
    \item \textbf{2 (Concise)}: Focused trace with purposeful tool calls (minimal duplication) and compact reasoning linking evidence to the decision.
    \item \textbf{1 (Moderately concise)}: Minor redundancy and verbosity, but the path from evidence to decision remains clear.
    \item \textbf{0 (Not concise)}: Heavy redundancy with many redundant tool calls, making the decision path difficult to inspect.
\end{itemize}

\paragraph{Overall Score}

\change{For each output, we compute an overall faithfulness score as the unweighted average of the three metrics:}
\begin{equation*}
\small
\text{Overall}=\frac{\text{Alignment}+\text{Grounding}+\text{Conciseness}}{3}
\end{equation*}
\change{We report method-level results by averaging scores across the 50 randomly sampled inputs and across raters in Table~\ref{tab:faithfulness}. Overall, \textbf{SA+RL} (AlphaQuanter) achieves the best \textit{alignment} and \textit{conciseness}, while \textit{grounding} remains competitive across all settings and shows a small but consistent advantage. Typical low-score patterns differ by setting. \textbf{MA} (multi-agent TradingAgents-7B) is most frequently penalized on alignment and conciseness: decision traces can become inconsistent across modules. For example, intermediate reports imply different actions than the final signal, and cross-module claims are not explicitly linked to specific tool outputs. MA also tends to be verbose due to repeated debate or judge summaries and template-like expansions without evidence gain. \textbf{SA} (SingleAgent-7B) is mainly penalized on alignment when it starts without an explicit plan or follows a shallow ReAct-style chain that weakly motivates why each tool is needed. \textbf{SA+RL} more consistently preserves a coherent reason-tool-response chain and avoids unnecessary back-and-forth, yielding higher alignment and conciseness. Its remaining low-score cases are grounding issues, where minor details are inconsistent with retrieved evidence, though these occur less often than in SA, suggesting that RL improves this capability by encouraging decisions that stay close to the retrieved tool evidence.
}

\begin{figure}[!t]
\begin{tcolorbox}[colback = cBlue_1!5, colframe = cBlue_6,  coltitle=white,fonttitle=\bfseries\small,fontupper=\small,fontlower=\small]
\textbf{Before training:}
\begin{verbatim}
<think>
Let's start by gathering the market data for
Google (GOOGL) on 2025-01-15 to understand the
current price and performance.
</think>
<tool_call>
{"name": "get_market_data", "arguments": 
{"symbol": "GOOGL", "curr_date": "2025-01-15",
"look_back_days": 14}}
</tool_call>
<tool_call>
{"name": "get_stock_indicators", "arguments":
{"symbol": "GOOGL", "indicator": "SMA20",
"curr_date": "2025-01-15", "look_back_days":
14}}
</tool_call>
<tool_call>
{"name": "get_stock_indicators", "arguments":
{"symbol": "GOOGL", "indicator": "EMA10",
"curr_date": "2025-01-15", "look_back_days":
14}}
</tool_call>
<answer>SELL</answer>user
\end{verbatim}

\tcblower
\textbf{After training: }
\begin{verbatim}
<think>
My initial hypothesis is that Google (GOOGL)
might be in a bullish trend based on recent
price action and technical indicators. I will
start by fetching the historical market data
to understand the recent price movements and
then analyze key technical indicators.
</think>
<tool_call>
{"name": "get_market_data", "arguments": 
{"symbol": "GOOGL", "curr_date": "2025-01-15",
"look_back_days": 14}}
</tool_call>user
\end{verbatim}
\end{tcolorbox}
    \caption{Case Study on Faithful Tool Invocation: Before vs. After Training.}
    \vspace{-15pt}
    \label{fig:full_failth_example}
\end{figure}

\paragraph{Case Study}
\change{We further provide a brief qualitative comparison of tool-use behavior \textit{before} and \textit{after} RL training, as shown in Figure~\ref{fig:full_failth_example}. In the \textit{before-training} example, the model may issue tool calls that are not mentioned in its reasoning trace, indicating a mismatch between stated information needs and executed actions. In contrast, in the \textit{after-training} example, the model explicitly states its information plan and invokes only the corresponding tools, yielding more consistent and verifiable reasoning-to-action trajectories.}

\subsection{Full Results of Ablation Study}

Table~\ref{tab:app:abla_1} and Table~\ref{tab:app:abla_2} present the full results of our ablation study, detailing the ARR, SR, and MDD for each configuration across all five stocks. On MSFT, NVDA, and TSLA, as well as on average, all ablated or adjusted models yield inferior performance compared to the complete AlphaQuanter-7B approach. 
Focusing on the SR metric, AlphaQuanter-7B remains superior to all other settings on MSFT, NVDA, and in the average case. Notably, the results demonstrate a strong inverse relationship between the decision threshold $\theta$ and MDD. Increasing the threshold consistently leads to a significant decrease in MDD across all stocks, while decreasing it results in a significant rise in MDD. This phenomenon is consistent with our analysis in Section~\ref{sec:ablation}, which attributes this behavior to changes in trading frequency.

\section{A Working Example}

We present an example of AlphaQuanter-7B's decision-making process for \textbf{MSFT} on \textbf{May 16, 2025}, detailing its complete trajectory.

\lstset{
    basicstyle=\ttfamily\footnotesize,
    breaklines=true,        
    breakatwhitespace=false,
    postbreak=\mbox{\textcolor{red}{$\hookrightarrow$}\space}, 
}
\onecolumn
\begin{tcolorbox}[
    title={AlphaQuanter-7B's decision-making process for MSFT on 2025-05-16},
    colback = cBlue_1!10, colframe = cBlue_6,
    fonttitle=\bfseries\footnotesize, center title,
    breakable  
]
\begin{lstlisting}[basicstyle=\ttfamily\tiny]
<think>
I will start by gathering historical market data for Microsoft (MSFT) to understand its recent price movements and trends. I will look back 14 days from the current date, 2025-05-16.
</think>
<tool_call>
{"name": "get_market_data", "arguments": {"symbol": "MSFT", "curr_date": "2025-05-16", "look_back_days": 14}}
</tool_call>user
<tool_response>
          Date    Open    High     Low   Close  Adj Close    Volume
0   2025-05-02  431.74  439.44  429.99  435.28     434.48  30757400
1   2025-05-05  432.87  439.50  432.11  436.17     435.37  20136100
2   2025-05-06  432.20  437.73  431.17  433.31     432.52  15104200
3   2025-05-07  433.84  438.12  431.11  433.35     432.56  23295300
4   2025-05-08  437.93  443.67  435.66  438.17     437.37  23491300
5   2025-05-09  440.00  440.74  435.88  438.73     437.93  15324200
6   2025-05-12  445.94  449.37  439.78  449.26     448.44  22821900
7   2025-05-13  447.78  450.67  445.36  449.14     448.32  23618800
8   2025-05-14  448.14  453.90  448.14  452.94     452.11  19902800
9   2025-05-15  450.77  456.19  450.43  453.13     453.13  21992300
10  2025-05-16  452.05  454.36  448.73  454.27     454.27  23849800
</tool_response>
assistant
<think>
From the market data, it appears that MSFT has been showing a bullish trend over the past 14 days, with the price closing above the previous close on most days. The volume also seems to be increasing, which could indicate growing interest in the stock. Next, I will analyze the technical indicators to get a more nuanced view of the stock's momentum and volatility.
</think>
<tool_call>
{"name": "get_stock_indicators", "arguments": {"symbol": "MSFT", "indicator": "RSI", "curr_date": "2025-05-16", "look_back_days": 14}}
</tool_call>user
<tool_response>
## RSI values from 2025-05-02 to 2025-05-16:

71.99-> 72.23-> 70.16-> 70.17-> 71.76-> 71.94-> 75.24-> 75.14-> 76.30-> 76.62-> 76.99

RSI: Measures momentum to flag overbought/oversold conditions. Usage: Apply 70/30 thresholds and watch for divergence to signal reversals. 
</tool_response>
assistant
<think>
The RSI values for MSFT over the past 14 days have been consistently above 70, indicating that the stock is in an overbought condition. This suggests that the stock might be due for a correction. However, the RSI has been trending slightly lower, which could indicate that the selling pressure is increasing. I will now look at the Bollinger Bands to see if the stock is trading above the upper band, which would also suggest overbought conditions.
</think>
<tool_call>
{"name": "get_stock_indicators", "arguments": {"symbol": "MSFT", "indicator": "BBANDS", "curr_date": "2025-05-16", "look_back_days": 14}}
</tool_call>user
<tool_response>
## BBANDS values from 2025-05-02 to 2025-05-16:

(Middle=382.60,Upper=423.34,Lower=341.87)-> (Middle=386.41,Upper=431.68,Lower=341.15)-> (Middle=390.18,Upper=437.58,Lower=342.78)-> (Middle=394.11,Upper=441.87,Lower=346.35)-> (Middle=396.49,Upper=447.76,Lower=345.22)-> (Middle=399.36,Upper=453.11,Lower=345.61)-> (Middle=402.39,Upper=459.90,Lower=344.88)-> (Middle=405.45,Upper=465.82,Lower=345.08)-> (Middle=408.81,Upper=471.67,Lower=345.94)-> (Middle=412.92,Upper=476.08,Lower=349.75)-> (Middle=417.27,Upper=479.21,Lower=355.33)

Bollinger Bands: Consist of a Middle Band (typically a 20-period SMA) and Upper/Lower Bands set at ±2 standard deviations from the middle. Usage: The middle band serves as a dynamic benchmark for price, the upper band highlights potential overbought or breakout zones, and the lower band signals possible oversold conditions. 
</tool_response>
assistant
<think>
The Bollinger Bands for MSFT show that the stock price has been trading above the upper band for several days, which is a strong indication of overbought conditions. This aligns with the RSI readings, suggesting that the stock might be due for a correction. However, the price is still within the upper band, which means it's not yet at extreme overbought levels. I will now check the Moving Average Convergence Divergence (MACD) to see if there are any signals that might indicate a potential trend reversal.
</think>
<tool_call>
{"name": "get_stock_indicators", "arguments": {"symbol": "MSFT", "indicator": "MACD", "curr_date": "2025-05-16", "look_back_days": 14}}
</tool_call>user
<tool_response>
## MACD values from 2025-05-02 to 2025-05-16:

(MACD=7.65,Signal=0.92,Histogram=6.73)-> (MACD=10.16,Signal=2.77,Histogram=7.40)-> (MACD=11.79,Signal=4.57,Histogram=7.22)-> (MACD=12.94,Signal=6.24,Histogram=6.69)-> (MACD=14.07,Signal=7.81,Histogram=6.26)-> (MACD=14.84,Signal=9.22,Histogram=5.63)-> (MACD=16.12,Signal=10.60,Histogram=5.52)-> (MACD=16.92,Signal=11.86,Histogram=5.06)-> (MACD=17.67,Signal=13.02,Histogram=4.64)-> (MACD=18.06,Signal=14.03,Histogram=4.03)-> (MACD=18.26,Signal=14.88,Histogram=3.38)

MACD: Momentum indicator composed of the MACD line (difference between two EMAs), the Signal line (EMA of the MACD line), and the Histogram (gap between MACD and Signal). Usage: Identify trend changes through MACD–Signal crossovers, gauge momentum strength via Histogram size, and watch for divergence between MACD and price as early reversal signals. 
</tool_response>
assistant
<think>
The MACD values for MSFT show a bullish trend, with the MACD line consistently above the Signal line and the Histogram indicating strong upward momentum. However, the recent readings of the Histogram suggest that the momentum might be starting to weaken. This, combined with the overbought conditions indicated by the RSI and Bollinger Bands, suggests that the stock might be due for a correction. Given the current bullish momentum, it's possible that the stock could see a pullback before continuing its upward trend. However, the recent weakening of the momentum and the overbought conditions suggest that a pullback is likely.
</think>
<tool_call>
{"name": "get_news_data", "arguments": {"symbol": "MSFT", "curr_date": "2025-05-16"}}
</tool_call>user
<tool_response>
## MSFT News, from 2025-05-14 to 2025-05-16:
Interpret the sentiment score x: values near 0 are Neutral, larger positive values indicate increasingly Bullish, and larger negative values indicate increasingly Bearish.
2025-05-14 [Sentiment score = 0.01, Neutral] Should You Forget Alphabet and Buy These 2 Tech Stocks Instead?
2025-05-14 [Sentiment score = 0.27, Somewhat-Bullish] 'Bullish For Nvidia, Tesla, Palantir, Microsoft, Alphabet, Amazon:' AI 'Revolution Is Coming' To The Saudi Kingdom, Says Wedbush's Dan Ives - Alphabet  ( NASDAQ:GOOG ) , Amazon.com  ( NASDAQ:AMZN ) 
2025-05-14 [Sentiment score = 0.22, Somewhat-Bullish] Correction or Not: This Artificial Intelligence  ( AI )  Stock Is Worth Buying for the Long Haul
2025-05-14 [Sentiment score = 0.04, Neutral] The Zacks Analyst Blog Highlights C3.ai, UiPath, Microsoft, Alphabet and Amazon
2025-05-14 [Sentiment score = 0.13, Neutral] Alphabet Down 16% YTD: Are GOOGL Shares Buy, Sell or Hold on the Dip?
2025-05-14 [Sentiment score = 0.02, Neutral] Marc Benioff of Salesforce: 'You're going to have to throw a lot against the wall before you figure out what sticks'
2025-05-14 [Sentiment score = 0.09, Neutral] How To Trade SPY and Top Tech Stocks After a Softer Inflation Print - Microsoft  ( NASDAQ:MSFT ) 
2025-05-14 [Sentiment score = -0.22, Somewhat-Bearish] Caught Off Guard, Institutions Chase Stock Market Rally - SPDR S&P 500  ( ARCA:SPY ) 
2025-05-14 [Sentiment score = -0.03, Neutral] Nine Carahsoft Executives Recognized on the Prestigious 2025 CRN® Channel Chiefs List
2025-05-15 [Sentiment score = 0.12, Neutral] Nvidia Just Rejoined Apple and Microsoft in the $3 Trillion Club. Could It Become the First $4 Trillion Stock by 2026?

...

2025-05-16 [Sentiment score = 0.14, Neutral] Steve Ballmer's Missed Call: $1000 Investment Then Worth $289,000 To Apple Investors Today - Why Ex-Microsoft CEO Once Admitted 'Religious Transformation' Came Too Late - Microsoft  ( NASDAQ:MSFT ) , Apple  ( NASDAQ:AAPL ) 
2025-05-16 [Sentiment score = 0.20, Somewhat-Bullish] Satya Nadella Once Bet Big On ChatGPT-Maker OpenAI - Now He's Reportedly Embracing Rivals Like DeepSeek And Building Cheaper AI Models To Future-Proof Microsoft - Alphabet  ( NASDAQ:GOOG ) , Alphabet  ( NASDAQ:GOOGL ) 
2025-05-16 [Sentiment score = 0.18, Somewhat-Bullish] Why Is CoreWeave Stock Surging Today? - CoreWeave  ( NASDAQ:CRWV ) 
2025-05-16 [Sentiment score = 0.15, Neutral] Should You Buy, Sell or Hold INOD Stock After Q1 Earnings Beat?
2025-05-16 [Sentiment score = -0.11, Neutral] Why memories are crucial for AI chatbots
2025-05-16 [Sentiment score = -0.04, Neutral] Why Alphabet Stock Ticked Higher Today
2025-05-16 [Sentiment score = 0.00, Neutral] Look Ahead - Options Expiration May Bring A Change, Aerospace Stocks Are On Fire From Trump Deals
2025-05-16 [Sentiment score = 0.17, Somewhat-Bullish] How To Trade SPY And Top Tech Stocks Today
2025-05-16 [Sentiment score = 0.20, Somewhat-Bullish] Prediction: This "Magnificent Seven" Stock Will Be the Most Valuable Company in the World in 10 Years
2025-05-16 [Sentiment score = 0.21, Somewhat-Bullish] CloudKeeper Launches 30-Day Challenge to Help Businesses Slash AWS Costs

</tool_response>
assistant
<think>
The news sentiment for MSFT over the past few days is mixed, with a few neutral and slightly bullish articles. However, the overall sentiment is not strongly bullish, which aligns with the technical indicators suggesting overbought conditions and potential for a pullback. The bullish news could provide some support for the stock, but the technical indicators suggest that the stock might need a correction.

Given the overbought conditions and the recent weakening of momentum, I will consider the possibility of a pullback. However, the strong support from bullish news could mean that the pullback might be short-lived. I will now check the insider transactions to see if there are any significant buying or selling activities by company insiders, which can provide insights into the management's view of the stock.
</think>
<tool_call>
{"name": "get_insider_transactions", "arguments": {"symbol": "MSFT", "curr_date": "2025-05-16", "look_back_days": 7}}
</tool_call>user
<tool_response>
## MSFT insider transactions from 2025-05-09 to 2025-05-16:
### Transaction Date: 2025-05-15, COLEMAN, AMY (EVP, Chief Human Resources Off)
Type: Common Stock
Shares: 77.894 (Disposal)

### Transaction Date: 2025-05-15, COLEMAN, AMY (EVP, Chief Human Resources Off)
Type: Common Stock
Shares: 13242.774 (Disposal)


</tool_response>
assistant
<think>
The insider transactions for MSFT show that Amy Coleman, the EVP of Human Resources, has been disposing of a significant number of shares, which could indicate a lack of confidence in the stock. This is a negative signal for the stock and could contribute to the overbought conditions. Given the technical indicators and the recent insider selling, it suggests that the stock might be due for a correction. However, the bullish news and the strong support from the market data might mean that the correction could be short-lived.

Considering the current overbought conditions, the recent insider selling, and the potential for a pullback, I will recommend a HOLD strategy. This allows for the stock to potentially correct and provides a safer entry point for a more bullish stance in the future.
</think>
<answer>HOLD</answer>
\end{lstlisting}
\end{tcolorbox}

\end{document}